\documentclass{emulateapj}

\newcommand{\dop}{\delta_{\rm D}}
\newcommand{\g }{\gamma}
\newcommand{\e }{\epsilon}

\begin{document}

\title{Statistics of Cosmological Black Hole Jet Sources:\\
Blazar Predictions for GLAST}

\author{Charles D. Dermer\altaffilmark{1}}

\altaffiltext{1}{E. O. Hulburt Center for Space Research, Code 7653
Naval Research Laboratory, Washington, D.C. 20375-5352; 
\small dermer@gamma.nrl.navy.mil}

\begin{abstract}
A study of the statistics of cosmological black-hole jet sources is
applied to EGRET blazar data, and predictions are made for GLAST.
Black-hole jet sources are modeled as collimated relativistic plasma outflows
with radiation beamed along the jet axis due to strong Doppler
boosting. The comoving rate density of blazar flares is assumed to 
follow a blazar formation rate
(BFR),  modeled by analytic functions based on astronomical
observations and fits to EGRET data. 
The redshift and size
distributions of gamma-ray blazars observed
with EGRET, separated into BL Lac object (BL) and flat spectrum radio
quasar (FSRQ) distributions, are fit with monoparametric functions for
the distributions of the jet Lorentz factor $\Gamma$, comoving
directional power $l_e^\prime$, and spectral slope.  
A BFR factor $\approx 10\times$ greater at  $z \gtrsim 1$ than at 
present is found to fit the FSRQ data.
A smaller comoving rate density and greater luminosity 
of BL flares at early times compared to the present epoch fits the 
BL data. Based on the EGRET
observations, $\approx 1000$ blazars consisting of $\approx 800$ FSRQs and FR2 
radio galaxies and $\approx 200$ BL Lacs and FR1 radio galaxies will be detected with GLAST
during the first year of the mission. Additional AGN classes, such as hard-spectrum
BL Lacs that were mostly missed with EGRET, could add more GLAST sources. The FSRQ and
BL contributions to the EGRET  $\gamma$-ray background at 1 GeV are
estimated at the level of $\approx 10$ -- 15\% 
and $\approx 2$\% -- 4\%, respectively.  EGRET and GLAST
sensitivities to blazar flares are considered in the optimal case, and
a GLAST analysis method for blazar detection is outlined.
\end{abstract}

\keywords{AGNs: blazars---black holes---gamma-ray bursts}

\section{\label{intro}Introduction}

Population studies of black-hole jet sources, which include blazars,
gamma-ray bursts, and microquasars, are difficult because of the
unknown emission processes and beaming patterns of the relativistic
jets.  Moreover, the density and luminosity evolution of black-hole
jet sources through cosmic time is uncertain. Here we develop a method
to treat the statistics of black-hole jet sources using the
$\gamma$-ray data alone. Although the focus of this study is radio
galaxies and blazars, the method can also be applied to GRBs \citep{ld06}.

The interest in population statistics of blazar sources is that an
accurate determination of source density evolution is needed to
identify parent populations \citep{up95}, to chart black-hole
formation and growth throughout the history of the universe
\citep{bd02,ce02,mt03}, and to assess the contribution of black-hole
jet sources to the $\gamma$-ray background. The isotropic $\gamma$-ray background 
\citep{sre98} consists of an extragalactic
$\gamma$-ray background (EGRB)  and an uncertain contribution of
quasi-isotropic Galactic $\gamma$ rays produced, for example, by
Compton-scattered radiations from cosmic-ray electrons \citep{smr00,smr04}.  

Soon after the recognition of the $\gamma$-ray blazar class with EGRET
\citep{fic94}, $\gamma$-ray blazar population studies were undertaken.
\citet{chi95} performed a $\langle V/V_{max}\rangle$ analysis assuming
no density evolution and showed that luminosity evolution of EGRET blazars 
was implied by the data. With a larger data set and using radio data to 
ensure the sample was unbiased in regard to redshift determination, 
\citet{cm98} again found that evolution was required. They
obtained best-fit values through the maximum likelihood method that
gave an AGN contribution to the EGRB at the level of
$\approx 25$\%. 

\citet{ss96} postulated a radio/$\gamma$-ray connection, and
tried to correct for the duty cycle and $\gamma$-ray spectral
hardening of flaring states.  They found that essentially 100\% of the
EGRET EGRB arises from unresolved blazars and AGNs.
They did not, however, fit the blazar redshift distribution to provide 
a check on their model.  The crucial underlying
assumption of this approach, which has been developed in further
detail in recent work \citep{gio06,nt06}, is that there
is a close connection between the radio and $\gamma$-ray properties of
blazars. Because a large number of EGRET blazars 
(FSRQs) are found
in the 5 GHz, $>1$ Jy \citet{kue81} catalog, a radio-$/\gamma$ ray
correlation is expected, but is not found in 2.7 and 5 GHz 
monitoring of EGRET $\gamma$-ray blazars \citep{mue97}.  X-ray
selected BLs are also not well-sampled in radio surveys. 
Studies based on correlations between the radio and $\gamma$-ray emissions
from blazars may therefore be based on a questionable foundation.  
It is therefore necessary to distinguish between the
very different properties and histories of FSRQs and BLs and their
separate contributions to the EGRB.

A detailed physical model to treat blazar statistics 
that avoids any radio/$\gamma$-ray blazar correlation 
 was developed by \citet{mp00}. 
Blazar spectra were calculated assuming an injection electron number index of $-2$.
Distributions in injected particle energy into BL and FSRQ jets were 
considered in the modeling. The indices in the injected energy 
distributions were taken from the luminosity functions of  
 FR1 and FR2 radio galaxies which, according to the blazar
unification scenario \citep{up95}, are the parent populations
of BLs and FSRQs, respectively. A simple description of density 
evolution is given in the form of a cutoff at some maximum 
redshift $z_{max}$. Depending on the value of $z_{max}$, 
\citet{mp00} concluded that as much as $\approx 40$
-- 80\% of the EGRB is produced by unresolved AGNs, with 
$\approx 70$ -- 90\% of the emission from FR1s and
BLs. 

Here we also treat a physical model for FSRQs and BLs. 
This is similar to the study of \citet{mp00}, 
though different in a number of important ways. 
No detailed radiation modeling is employed, but
rather we use mean spectral indices as measured by EGRET. 
Various  model forms  
describe the rate density of blazar flares, but 
only single values of luminosity and $\Gamma$ factor 
are considered for each of the FSRQ and BL classes.
A mono-luminosity function for 
blazars means that the range in apparent powers is kinematic, 
arising from the different, randomly
oriented jet directions.  This simple blazar model is highly constrained when fitting
to data, even given the freedom to consider different redshift-dependent
analytic forms for the blazar formation rate (BFR).  In the case of detailed fits to 
blazar data,  the assumption of no luminosity evolution can be and, for BLs, is 
relaxed.

We use parameter sets that give acceptable agreement 
to the  EGRET  data on blazar redshift 
and size distributions  to make predictions
that will be tested with GLAST,\footnote{glast.gsfc.nasa.gov, www-glast.stanford.edu}
and to estimate the blazar contribution to the EGRB.
The connection of the
properties of $\gamma$-ray blazars to blazars detected at radio, X-ray, and
other wavelengths can be used to determine the accuracy of  
models that assume a radio/$\gamma$-ray connection.

Section 2 describes the sample of EGRET blazars used.
The equations for the analysis are presented in Section 3, and results
of the parameter study are described in Section 4. Predictions for
GLAST and estimates for the EGRB from 
unresolved blazars are presented in Section 5, and we summarize in Section
6. Sensitivities of EGRET and GLAST to blazar flares, expressions for
optimal sensitivities, and a discussion of a GLAST analysis strategy
for blazar populations are given in the Appendix A. The self-absorption 
frequency is derived in Appendix B.

\section{Sample}

Crucial to making a proper comparison of a model to EGRET blazar data
is to choose a sample that is unbiased with respect to exposure and
background. For example, exposure to the region around 3C 273 and 3C
279 was much longer than average over the lifetime of EGRET, so
blazars found in these pointings would be detected to much smaller
flux thresholds than on average. Likewise, blazars in the vicinity of
the galactic plane would have to be much brighter than high-latitude
sources to be detected above the diffuse galactic $\gamma$-ray
emission.

The 18 month all-sky EGRET survey \citep{fic94}, which ran from 1991
May to 1992 November, had roughly uniform exposure over all parts of
the sky. Thirty-eight AGN identifications were reported in this
catalog. Additional analysis of the Phase 1 data, as reported in the
3EG catalog \citep{hea99}, revealed numerous additional detections of
AGNs during Phase 1.

We have identified all the high-confidence blazars listed in the Third
EGRET catalog that also appear during the 18 month all-sky survey,
during which all parts of the sky receive roughly uniform exposure.
The sample we use consists of 60 high-confidence gamma-ray blazars,
consisting of 14 BLs and 46 FSRQs.  We exclude sources within
10$^\circ$ of the Galactic plane, and use source catalogs
\citep{pg95,pea96} to establish BL identifications. The integral 
photon number fluxes $\phi(>E)$, in units of ph($>$100 MeV) cm$^{-2}$ 
s$^{-1}$, were
used to construct the FSRQ and BL size distributions.

\begin{figure}[t]
\epsscale{1.0}
\vskip-1.0in
\plotone{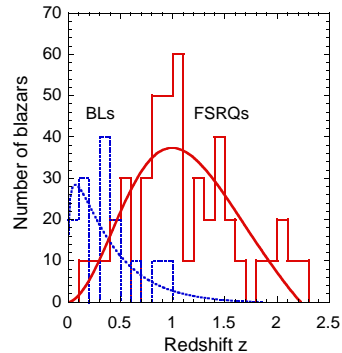}
\vskip-1.0in
\caption{EGRET observations of the redshift distributions of blazars,
separated into FSRQ (solid) and BL Lac (dotted) populations, are shown
by the histograms. Smooth solid curve shows a monoparametric FSRQ
blazar fit with BFR IR,4 (see Fig.\ 3), $\Gamma = 10$, $p= 3.4$, EC statistics,
and $l_e^\prime = 2.5\times 10^{39}$ ergs s$^{-1}$ sr$^{-1}$. The
dotted curve shows the monoparametric blazar fit for the BL data for a
blazar model with  $\Gamma = 4$, $p = 3.0$,
$l_e^\prime (z) = 6\times 10^{42}z^{1.95}$ ergs s$^{-1}$ sr$^{-1}$, synchrotron/SSC
statistics, and $\dot n_{BL}(z)
\propto z^{-9/4}$. }
\label{fig1}
\end{figure}

Table 1 lists the sources from the Third EGRET catalog \citep{hea99}
used in this study and their classifications. The $\gamma$-ray blazar
sample, binned by redshift and grouped into BLs and FSRQs, is plotted
in Fig.\ \ref{fig1}. Fig.\ \ref{fig2} shows the FSRQ and BL size
distributions.  As shown in Appendix A, 
the on-axis EGRET sensitivity for a two-week
pointing to high-latitude blazars is $
10^{-8}\phi_{-8}$ ph($>100$ MeV) cm$^{-2}$ s$^{-1}$, with
$\phi_{-8} \cong 15$. This is in accord with the
dimmest blazars detected with EGRET during the 
two-week pointings. The corresponding $\nu F_\nu$ 
threshold flux is 
$f^{thr}_\e = E^2\phi_s(E)$, where the source flux is given 
by eq.\ (\ref{phise}). Thus 
\begin{equation}
f^{thr}_\e \cong 1.6\times 10^{-12}(\alpha_{ph} - 1) \phi_{-8} {\rm ~ergs~cm}^{-2}~s^{-1}\;
\label{fthr}
\end{equation} 
for measurements at $E>$ 100 MeV ($\e \gtrsim 200$), 
where $\alpha_{ph}$ is the number spectral index. For $\phi_{-8} \cong 15$,
$f^{thr}_\e \cong 2.4\times 10^{-11}(\alpha_{ph} - 1)$ ergs cm$^{-2}$ s$^{-1}$, 
noting that $\alpha_{ph} \cong (p+1)/2$ for Thomson, synchrotron, 
and SSC processes, where $p$ is the number index of the assumed power-law 
electron distribution. 
As shown in Appendix A, an energy range that
depends on the source spectral hardness should be used
for optimal detection significance of blazars 
with GLAST or EGRET.

\begin{figure}[t]
\epsscale{1.0}
\vskip-1.5in
\plotone{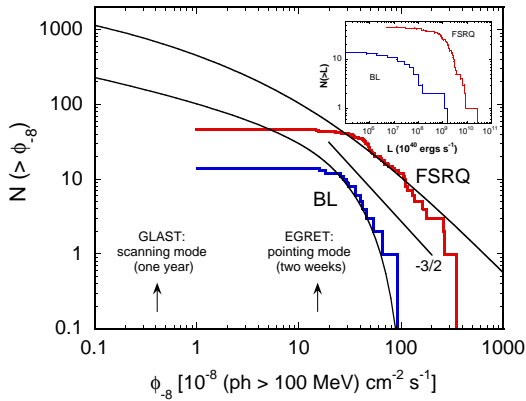}
\caption{EGRET observations of the size distribution of blazars 
in terms of the integral photon fluxes at photon energies
$E\gtrsim 100$ MeV versus model size distributions for
FSRQ and BL blazars, with parameters given in the caption to Fig.\ 1.
Also shown are the integral flux sensitivities for EGRET in the pointing mode 
for a two-week observation (over $\approx 1/24^{th}$ of the full sky) and for
GLAST in the scanning mode for one year (over the full sky). The line labeled
``$-3/2$" has the slope of the size distribution of  monoluminous sources uniformly 
distributed in flat space. The inset shows the size 
distribution of EGRET blazars in terms of apparent blazar luminosities in the 
energy range 100 MeV  -- 5 GeV. }
\label{fig2}
\end{figure}

The BLs and FSRQs display very different $\gamma$-ray properties. The
BLs are less numerous and closer, with an average redshift $\langle z\rangle \sim
0.37$. For the FSRQs,  $\langle z\rangle \sim 1.11$, with a tail 
on the FSRQ distribution 
reaching to $z \sim 2.3$.  There are $\approx 5\times$ as many
FSRQs as BLs per unit peak flux in the EGRET range $\phi_{-8}\approx 25$
-- 100.  The FSRQs are $\sim 1$ -- 2 orders of
magnitude more luminous than the BLs (inset to Fig.\
\ref{fig2}). The apparent powers of the 
FSRQs are as large as $\approx
10^{50}$ ergs s$^{-1}$, compared to BLs, which are
 typically $\lesssim 
10^{48}$ ergs s$^{-1}$.

More detailed studies of the redshift and size distribution can be
made with the EGRET data (R.\ Romani, private communication, 2006)
based on surveys to identify $\gamma$-ray blazars \citep{sow05}. These
studies also show that the mean redshift of FSRQs is $\gtrsim 1$, and
that both the BL and FSRQ populations have tails to high
redshifts. The highest redshift $\gamma$-ray blazar candidate has $z =
5.47$ \citep{rom04}.  Thus we can be sure that the FSRQ BFR extends
to at least $z \gtrsim 5$.

The Third EGRET catalog gives the two-week average fluxes for 
blazars during Phase 1. In comparison with models, the flaring
timescale of the blazars is given by a rate density, which relates
to the source density by a flaring timescale and duty cycle factor.
We take this flaring timescale to be $\approx 2$ weeks for the observer, 
and set the duty cycle equal to unity. If the duty cycle is less than 
unity, then the source space density must be greater, leaving the
calculation of the intensity from unresolved sources independent
of these factors.

\section{Analysis}

We employ a simplified version of the standard model for blazars
considered by \citet{dg95} \citep[see also][]{sik97,dd00}. A
relativistically moving plasmoid ejected from a black-hole engine has
accelerated within it, either through internal or external shocks or
otherwise, a power-law distribution of quasi-isotropic
ultra-relativistic electrons with number index $p$.  In its proper
frame, the plasmoid is assumed to entrain a randomly oriented magnetic
field with mean strength $B$.  The nonthermal electrons emit
synchrotron or Thomson radiation that is Doppler boosted by the
effects of the relativistic motion.  For the calculations shown here,
we use the synchrotron beaming factor $\propto \dop^{(5+p)/2}$ for 
synchrotron and synchrotron self-Compton (SSC) processes, where
the Doppler factor $\dop = [\Gamma(1-\beta\cos\theta)]^{-1}$, and the
external Compton (EC) beaming factor $\propto \dop^{3+p}$
\citep{der95}, which also holds for scattering in the Klein-Nishina
regime \citep{gkm01}.  Here $\Gamma$ is the bulk Lorentz factor of the
plasma blob, $\beta = \sqrt{1-\Gamma^{-2}}$, $\alpha = (p-1)/2 =
\alpha_{ph}-1$ is the energy spectral index of the radiation,
and $\theta$ is the angle between
the jet and line-of-sight directions.  Continuous outflow scenarios
produce a beaming pattern weaker by one power \citep{lb85}, but the
single plasmoid approximation is more applicable to the flaring blazars.

\subsection{Cosmology of Nonthermal Sources}

Consider a power-law distribution of electrons with low- and
high-energy cutoffs, so that the total number distribution of
nonthermal electrons within the plasmoid is described by
\begin{equation}
N_e^\prime(\g ) = K^\prime_e \g^{-p}\;H(\g;\g_1,\g_2)\;.
\label{Neg}
\end{equation}
Note that $p$ is the number index of the emitting electrons, and could
be different from the injection index if cooling is important.
Normalizing to the total comoving electron energy $W_e^\prime =
m_ec^2\int_1^\infty d\g\;\g\;N_e^\prime(\g )$ implies
$$K^\prime_e = 
{(p-2) W_e^\prime\over m_ec^2}\;(\g_1^{2-p} - \g_2^{2-p})^{-1}
\rightarrow$$
\begin{equation}
 {(p-2)W_e^\prime \g_1^{p-2}\over m_ec^2}\;,\;{\rm when~}
 p>2\;,\;\g_2\gg\g_1\;.
\label{Ke}
\end{equation}
Cooling can introduce a break in the electron spectrum \citep{mp00},
but the EGRET spectra of bright blazars are well fit by a single
power law \citep{muk97}, indicating that the single power-law approximation is
adequate for a treatment of blazar statistics.

The $\nu F_\nu$ nonthermal synchrotron radiation spectrum for a
comoving isotropic power-law distribution of electrons entrained in a
randomly oriented magnetic field is given in the $\delta$-function
approximation by the expression
\begin{equation}
f^{syn}_\e \cong {\dop^4 \over 6\pi d_L^2}\;c\sigma_{\rm T} U_B\g_s^3 
N_e^\prime (\g_s )\;,\;\g_s = \sqrt{{\e_z\over \dop b}}\;,
\label{fes}
\end{equation}
where the luminosity distance
for a flat $\Lambda$CDM universe is
\begin{equation}
d_L(z) = {c\over H_0}
\;(1+z)\int_0^z dz^\prime\;{1\over \sqrt{\Omega_m (1+z^\prime)^3 +
\Omega_\Lambda}}\;,
\label{dlum}
\end{equation}
$U_B = B^2/8\pi$ is the magnetic-field energy density in the jet
plasma, $b \equiv {B/ B_{cr}}$, and $B_{cr} = {m^2_e c^3/e\hbar}$ is
the critical magnetic field.

The $\nu F_\nu$ spectrum of jet electrons that Thomson scatter an
external quasi-isotropic monochromatic radiation field with stationary
(explosion)-frame dimensionless photon energy $\e_* = 10^{-4} \e_{-4}$
and stationary frame energy density $U_*$ is
\begin{equation}
f^{EC}_\e \cong {\dop^6 \over 6\pi d_L^2}\;c\sigma_{\rm T} 
U_*\g_{\rm C}^3 N_e^\prime (\g_{\rm C} )\;,
\;\g_{\rm C} = {1\over \dop}
\sqrt{{\e_z\over 2\bar\e_*}}\;
\label{fec}
\end{equation}
\citep[\citet{der95}; see][for expressions describing the Thomson-scattered
accretion-disk radiation fields]{ds02}. Restriction to the Thomson
regime implies that $\e_z \lesssim 1/(8\e_*)$, so that a target 5 eV
UV radiation field would display effects from the onset of the KN
decline in the cross section at $E \gtrsim 6$ GeV$/(1+z)$. A more
accurate treatment for GLAST analyses will have to consider the
effects of the KN decline on the statistics.

The $\nu F_\nu$ synchrotron self-Compton (SSC) radiation spectrum in
the $\delta$-function approximation is
\begin{equation}
f^{SSC}_\e \cong {\dop^4 \over 9\pi d_L^2}\;
{c\sigma_{\rm T} r_b U_B K_e^{\prime~2}\over 
V_b^\prime}\;
\g_s^{3-p} \;\Sigma_{\rm C}\;,
\label{fessc}
\end{equation}
where the Compton-synchrotron logarithm $\Sigma_c =
\ln(a_{max}/a_{min})$, $a_{max} = \min(b\g_2^2,
\e^\prime/\g_1^2,\e^{\prime ~-1})$, $a_{min} =
\max(b\g_1^2,\e^\prime/\g_2^2)$, and $\e^\prime = \e_z/\dop$
\citep{gou79,dss97}. 
The SSC process has a similar dependence as the synchrotron
process---though a curvature in the spectrum is produced by
$\Sigma_c$---but with a different coefficient that depends on the
physical size of the radiating plasma.

In this formulation, the radiating plasma is spherical in the comoving
frame, with volume $V_b^\prime = 4\pi r_b^{\prime~3}/3$, where the 
blob radius
\begin{equation}
r_b^\prime = {c t_{var}\delta_{\rm D}\over 1+z} \lesssim {2\Gamma c t_{var}\over 1+z}\;,
\label{rbprime}
\end{equation}
and $t_{var} ({\rm s}) = 86400t({\rm day}) = 10^3 t_3$ is the measured variability time scale. 
For FSRQs measured with EGRET, 
$$r_b^\prime \lesssim {2.6\times 10^{16}(\Gamma/10) t({\rm day})\over (1+z)/2}\;{\rm cm} 
\simeq 0.01 t({\rm day}) (\Gamma/10)\;{\rm pc}\;,$$
and for BLs measured with EGRET
$$r_b^\prime \lesssim 2.4\times 10^{14}(\Gamma/4) t_3\;{\rm cm} 
\simeq 10^{-4} t_3 (\Gamma/4)\;{\rm pc}\;.$$
For these blob sizes, the synchrotron radiation at GHz frequencies is likely to 
be heavily self-absorbed, as shown in Appendix B. This calls
further into question the use of any radio/$\gamma$-ray correlation.

We rewrite eqs.\ (\ref{fes}) -- (\ref{fessc}) as 
\begin{equation}
f_\e^{proc} = {l^\prime_{e}\over d_L^2} \;\dop^q
\;\e_z^{\alpha_\nu}\;,
\label{feproc1}
\end{equation}
where $\alpha_\nu = (3-p)/2$ is the $\nu F_\nu$ spectral index, the
directional comoving luminosity 
\begin{equation} l^\prime_{e}({\rm ergs~s}^{-1}
{\rm sr}^{-1}) = {K_e^\prime c \sigma_{\rm T}\over 6\pi}
\cases{U_{B_{cr}} b^{(p+1)/2}, & syn $~$ \cr\cr {K_e^\prime
\sigma_{\rm T} \over 2\pi r_b^{\prime ~2}}
U_{B_{cr}}\Sigma_{\rm C}\;b^{(p+1)/2}, & SSC $~$ \cr\cr
U_*(2\bar\e_*)^{(p-3)/2}, &EC$~$ \cr}
\label{leprime5}
\end{equation}
the beaming factor index 
$$ q = \cases{(p+5)/2,\; & synchrotron, SSC $~$ \cr\cr p+3\;,
&EC$~$ \cr}\;,\; $$ 
and $U_{B_{cr}} \equiv B_{cr}^2/8\pi$.

The event rate per sr (or the directional event rate) 
for bursting sources in a $\Lambda$CDM cosmology 
 is
\begin{equation}
{d\dot N\over d\Omega} = {c\over H_0} \int_0^\infty dz\; {d^2_L(z)\;
\dot n_{com}(z) \over
(1+z)^3\;\sqrt{\Omega_m(1+z)^3 +\Omega_\Lambda}}\;,
\label{ddotNdO}
\end{equation}
where $\dot n_{com}(z)$ is the rate
density of sources at redshift $z$ \citep[see][for a detailed derivation]{der06}. 
From the WMAP data \citep{spe03},
we take $\Omega_m = 0.27$, $\Omega_\Lambda = 0.73$, and Hubble's
constant $H_0 = 72$ km s$^{-1}$ Mpc$^{-1}$.  The directional event
rate (i.e., number count) of steady sources is
\begin{equation}
{dN\over d\Omega} = {c\over H_0} \int_0^\infty dz\;{d^2_L(z)\; n_{com}(z)
 \over
(1+z)^2\;\sqrt{\Omega_m(1+z)^3 +\Omega_\Lambda}}\;, 
\label{dNdO}
\end{equation}
and $ n_{com}(z)$ is the differential source density.

In an integral formulation for bursting sources \citep{der92}, the
observed directional event rate above the $\nu F_\nu$ spectral flux
threshold $f^{thr}_\e$ of the telescope is 
$${d \dot N(>
f^{thr}_\e )\over d\Omega } = {c\over H_0}
\;\int_{f^{thr}_\e}^\infty d f_\e \; \int_0^\infty dl_e^\prime\;
\int_{-\infty}^{\infty} dp\; 
\int_1^\infty d\Gamma\;\times $$
\begin{equation}
 \int_{-1}^1\;d\mu \; \int_0^\infty dz\;{d^2_L(z)\; \dot n_{com}
(l_e^\prime,p,\Gamma; z) \over
 (1+z)^3\;\sqrt{\Omega_m(1+z)^3 +\Omega_\Lambda}}\;\times $$
$$  \delta[f_\e -
 f_\e^{proc} (l_e^\prime,p,\Gamma,\mu)]\;.
\label{N>f}
\end{equation}
Here $f_\e^{proc}$ is the $\nu F_\nu$ flux for the process under
consideration, with synchrotron, EC, and SSC fluxes written as eq.\
(\ref{feproc1}), $\mu = \cos \theta$, and the blazar jet luminosity is 
charaterized by $l_e^\prime$.  For a better treatment of detector response, one
should calculate a photon-energy integration over effective area,
rather than describing a $\gamma$-ray telescope by a $\nu F_\nu$ flux
sensitivity $f^{thr}_\e$ at a single photon energy $\e$. 

For mono-parameter $\delta$-function distributions of $p$, $\Gamma$,
and $l_e^\prime$, we have
$$ {d\dot N(>f^{thr}_\e )\over d\Omega}  = {2 c\over H_0}
 \;
\int_0^\infty dz\;{d_L^2(z)\; \dot n_{com}(z)  \over
(1+z)^3\;\sqrt{\Omega_m(1+z)^3 +\Omega_\Lambda}}\;
 $$
\begin{equation}
\times \int_{-1}^1\;d\mu \;\int_{f^{thr}_\e}^\infty d f_\e\; \delta[f_\e -
f_\e^{proc} (l_e^\prime,p,\Gamma,\mu)]\;,
\label{N>f_1}
\end{equation}
where the various parameters specifying emission properties, for example,
$W^\prime_e$, B, $r_b^\prime$, and $U_*$, are found in the directional power $l_e^\prime$.
In this expression, we include a factor of 2 for a two-sided jet.

\subsection{Peak Flux and Size Distribution}

It is trivial to perform the integration over
$df_\e$ in eq.\ (\ref{N>f_1}), which places a limit
on the allowed values of $\mu$.
Only values of cosine angle $\mu \geq \hat\mu$ give 
detectable fluxes, where
\begin{equation}
\hat\mu(z,\Gamma,l_e^\prime,p,q,\e,f_\e) = \hat\mu\;=\; {1\over \beta}\; 
\big[ \; 1 - {1\over \Gamma}\big( {l_e^\prime\e_z^{\alpha_\nu}
\over d_L^2 f_\e}\big)^{1/q} \;\big]\;.
\label{hatmu}
\end{equation}
The $\nu F_\nu$ flux size distribution of blazars per sr per s is
therefore given by
\begin{equation}
{d\dot N \over d\Omega}(>f_\e ) \; = \;{ 2c\over H_0 }\;\int_0^\infty
dz\; {d_L^2(z)\; \dot n_{com}(z)[1 - \max(-1,\hat\mu)] \over (1+z)^3\;\sqrt{\Omega_m(1+z)^3
+\Omega_\Lambda} }
\label{ddotNdz_1}
\end{equation}
for two-sided blazar jet sources.

\subsection{Redshift Distribution}

The directional redshift distribution of sources with 
 $\nu F_\nu$ flux $ f_\e > f^{thr}_\e$
is simply given by
$${d\dot N \over dz d\Omega}(>f^{thr}_\e ) \; = $$
\begin{equation}
\;{ 2c\over H_0 }\;{d_L^2(z)\;
\dot n_{com}(z) \over (1+z)^3\;\sqrt{\Omega_m(1+z)^3 +\Omega_\Lambda}
}\;[1 - \max(-1,\hat\mu)]\;.
\label{ddotNdz}
\end{equation}
The threshold limitation prescribed by $\hat\mu$ in eq.\ (\ref{hatmu}) 
now has $f_\e \rightarrow f_\e^{thr}$. The relation between $f_\e^{thr}$
and the integral photon number flux variable $\phi_{-8}$ depends on $p$ and $\alpha_\nu$
as given by eq.\ ({\ref{fthr}).

\subsection{Parameters}

Eq.\ (\ref{feproc1}) shows that a $\gamma$-ray blazar can be detected from redshifts of order unity at
the level $f_\e $ when the parameters satisfy the condition
\begin{equation}
f_\e \cong \; 
l_e^\prime {\Gamma^{q} (2 \e)^{\alpha_\nu}\over d_L^2(z=1)}\;
\label{febnd}
\end{equation} 
which holds when $\Gamma \gg 1$.  If a significant fraction of the
EGRB originates from blazars, then $p \cong 3.2$. For synchrotron, SSC,  or
Thomson processes, this implies a photon number index $\alpha_{ph} =
2.1$, or $\alpha_\nu = -0.1$.  EGRET observations of blazar spectral
indices show that $\alpha_{ph} = 2.03\pm 0.09$ for BLs and
$\alpha_{ph} = 2.20\pm 0.05$ for FSRQs, with an average spectral index
of $\alpha_{ph} = 2.15\pm 0.04$
\citep{muk97}.  The evidence from EGRET that $\alpha_{ph}$ is
harder for BLs than for FSRQs means that the $\nu F_\nu$ 
peaks of the $\gamma$-ray components of BLs are typically 
at higher energies than for FSRQs.  Moreover,
observations suggest that the flaring state spectra are harder than
the quiescent emission, at least in the case of PKS 0528+134
\citep{muk96}.

Taking $\alpha_\nu = 0$ ($p = 3$) and noting that $d_L(z = 1) \cong 2.0\times 10^{28}$ cm, 
eq.\ (\ref{febnd}) becomes
\begin{equation}
f_\e \cong   2.5\times 10^{-11}\big({\Gamma\over 10}\big)^6 
\left({ l_e^\prime \over 10^{40}{\rm 
~ergs~s^{-1}~sr^{-1}}}\right ){\rm ergs~cm^{-2}s^{-1}}\;
\label{febndE}
\end{equation}
for EC statistics.
For synchrotron or SSC statistics, 
\begin{equation}
f_\e \cong   2.5\times 10^{-11}\big({\Gamma\over 10}\big)^4 
\left({ l_e^\prime \over 10^{42}
{\rm ~ergs~s^{-1}~sr^{-1}}}\right){\rm ergs~cm^{-2}s^{-1}},
\label{febndE1}
\end{equation}
so that a larger internal synchrotron power compared to 
Compton power is needed for the same $\nu F_\nu$ flux
when viewing within the beam of the jet.

In actual fitting of blazar statistical distributions, a degeneracy
between $\Gamma$ and $l^\prime_e$ is found that may partially be removed by
obtaining limits on the bulk Lorentz factor $\Gamma$ from
$\gamma\gamma$ attenuation arguments \citep[e.g.][]{mgc92,dg95}.

For standard FSRQ blazar parameters, we take 
\begin{equation}
p = 3.4\;,\;\Gamma = 10\;,\;l_e^\prime = 10^{40}
{\rm~ergs~s}^{-1} {\rm~sr}^{-1},\;
\label{standardfsrq}
\end{equation}
and EC statistics. For standard BL blazar parameters, we take 
\begin{equation}
p = 3.0\;,\;\Gamma = 4\;,\;l_e^\prime = 10^{42}
{\rm~ergs~s}^{-1} {\rm~sr}^{-1},
\label{standardbl}
\end{equation}
 and syn/SSC statistics.

For two week ($\approx 10^6$ s) observations,
$\approx 50$ flaring blazars were detected by EGRET
above its $\nu F_\nu$ threshold flux $f_\e^{thr}\cong 2.4\times 10^{-11}$ ergs
cm$^{-1}$ s$^{-1}$  over $\approx 1$ year. Given the EGRET field-of-view (see Fig.\ 2), 
this implies
a directional blazar flaring rate $\approx 4$ sr$^{-1}$ Ms$^{-1}$, implying
a fiducial rate density $\dot n_{com}(z = 1)$ at redshift unity of
about
\begin{equation}
4\times 10^{-6}{\rm~s}^{-1}{\rm sr}^{-1} \simeq { c\over H_0}\; 4\times 10^{56} {\rm cm}^2\;
\dot n_{com}(z = 1)\;,
\label{dotncoset}
\end{equation}
so that $\dot n_{com}(z = 1) \cong 0.8\times 10^{-90}$ events cm$^{-3}$
s$^{-1}$.  Thus we expect that the blazar flaring rate 
density $\dot n_{com} = 10^{-90}$ events cm$^{-3}$ s$^{-1}\approx$ 1 event Gpc$^{-3}$ yr$^{-1}$.
 The calculations for the local rate density of FSRQs give values larger
by one or two orders of magnitude. The discrepancy can be resolved
by considering the increased number of sources necessary to 
offset the reduction in the detection rate by $\approx \Gamma^{-2}$ for 
the beamed blazar emission, and the different rate density at $z<<1$
compared to $z = 1$, which depends on the form of the BFR.

\subsection{Blazar Formation Rate Histories}

\begin{figure}[t]
\epsscale{1.0}
\plotone{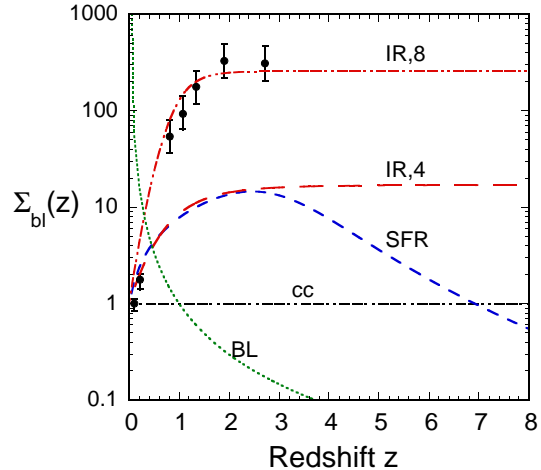}
\caption{Model blazar formation rates (BFRs) used in this study. 
The curve labeled ``cc" is the constant comoving rate, the curve 
``SFR" is the star formation rate from \citet{hb06}, 
curve ``IR,8" is eq.\ (\ref{bfr4}) with $n = 8$ that is used to fit the 
IR luminosity density from IR luminous galaxies, curve 
``IR,4" is  eq.\ (\ref{bfr4}) with $n = 4$ used to fit FSRQ data, and curve ``BL"
is  eq.\ (\ref{nBLz}) with $a = 1.75$ used to fit to the BL data.}
\label{fig3}
\end{figure}

The BFR functions are used to describe the change in the rate density
of blazars through cosmic time. This does not mean that the 
comoving density of supermassive black holes changes, but rather
that their flaring rate changes or their assignment to the FSRQ or BL class 
 changes with time as a consequnce of their definitions in terms 
of the optical emission line equivalent widths.

We consider the following BFRs $\Sigma_{bl}(z)$, shown in 
Fig.\ \ref{fig3},
which give the blazar
comoving rate density $\dot n_{com}(z) =  \Sigma_{bl}(z) \dot n_{com}$ in 
terms of the local comoving rate density $\dot n_{com}$:
\begin{enumerate}
\item  Constant comoving rate density, so that $\Sigma_{cc}(z) = 1$. 
This form is employed for mathematical convenience.
\item Comoving rate  $\propto $ the blue/UV luminosity density, 
which is assumed to track the star formation rate (SFR) of the
universe.  We use the analytic form \citep{hb06}
\begin{equation}
\Sigma_{SFR}(z) = {1+6.78z \over 1 + (z/3.3)^{5.2}}\;.\;
\label{SFR}
\end{equation}
Blazar activity could
be related to the SFR if stellar activity provides fuel for the
supermassive black hole engine, for example, from material driven off
by starburst nurseries encircling the nucleus \citep[e.g.,][]{hco04}.
\item Comoving rate $\propto $ sub-mm/far-IR luminosity density 
associated with luminous IR galaxies \citep{san04}, which we 
fit using the analytic form
\begin{equation} 
\Sigma_{IR,n}(z)= \;{1+ 2^{-n} \over (1+z)^{-n} + 2^{-n}}\;\;.
\label{bfr4}
\end{equation}
We obtain a good fit to the data with $n = 8$, as shown in 
Fig.\ 3. 
If IR-luminous galaxies are caused by galactic merger events, as
is indicated by morphological and spectral evidence 
\citep[e.g.,][and references therein]{sm96}, this would 
connect blazars and the formation of supermassive black holes to
galaxy collisions. Although related
to supermassive black hole growth, the IR luminosity density does not,
however, directly measure the activity of supermassive black holes, because
the IR radiation is a convolution of the photon luminosity 
which is then reprocessed through thick columns of material. 
Hence we have generalized the form with a single adjustable parameter, $n$,
that represents a range of BFR histories. The forms of $\Sigma_{IR,8}$ and 
$\Sigma_{IR,4}$ are shown in Fig.\ 3.
\item A BFR where the blazar flare rate density increases with 
cosmic time, which is found necessary to fit the BL data. The simple
form considered is 
\begin{equation}
\dot n_{BL}(z) = {\dot n_{BL}(z=1)\over z^a}\;,
\label{nBLz}
\end{equation}
where $a(>0)$ is adjusted to fit the data. Because of the divergence in the 
rate density when $z \rightarrow 0$, this BFR has to be normalized at $z>0$.
The total blazar flaring rate is, however, a well-defined value when $a <3$.
\end{enumerate}

\subsection{Diffuse Intensity}
 
The apparently diffuse intensity from the superposition of emissions
from many faint, unresolved blazars is given by \citep[e.g.,][]{der06}
\begin{equation}
\e I_\e = {2c\over 4\pi H_0}\;\int_0^\infty 
dz\;\oint d\Omega^\prime\; {\e_*^2\;q_{com}(\e_*,\Omega^\prime;z)\over
(1+z)^2 \;\sqrt{\Omega_m(1+z)^3 +\Omega_\Lambda}}\;.
\label{eIe}
\end{equation}
where $\e_* = \e_z = (1+z)\e$
and a factor of 2 is again introduced for two-sided
jet sources. The direction vector $\Omega^\prime$ defines the direction 
of the jet axis with respect to the observer direction.
This expression applies to persistent blazar sources,
with comoving emissivity 
\begin{equation}
\e_*^2 q_{com}(\e_*,\Omega^\prime;z) =
{d{\cal E}_\gamma\over dt_* dV_{com} d\Omega^\prime }\;= 
n_{com}(z)l_{e}^\prime\dop^q\e_z^{\alpha_\nu}
\;.
\label{qco}
\end{equation}
When sources are detected above threshold flux $f^{thr}_\e$, they no longer contribute
to the EGRB, as they are identified as a blazar source. 
This restricts eq.\ (\ref{eIe}) only to those blazars
with $ l_{e}^\prime\dop^q\e_z^{\alpha_\nu}/d_L^2 < f^{thr}_\e$, leading
to the following expression for the diffuse radiation from unresolved
radio galaxies and black hole jet sources:
$$\e I_\e^{bl}(<f_\e) \cong {c\e^{\alpha_\nu}
\over (q-1) H_0 \beta\Gamma^q}\int_0^\infty dz {(1+z)^{\alpha_\nu -2} \over 
\sqrt{\Omega_m(1+z)^3 + \Omega_\Lambda}}\times
$$
\begin{equation}
\;n_{com}(z)l_{e}^\prime(z)\big\{[1-\beta\min(1,\hat\mu)]^{1-q} - (1+\beta )^{1-q}\big\}\;.
\label{eIebl}
\end{equation}
The dependence of this expression on $f^{thr}_\e$ is carried
by $\hat\mu$, given by eq.\ (\ref{hatmu}).

Equating the EGRET two-week
average fluxes with blazars that flare once in two weeks
allows us to replace the comoving density of blazar AGN sources,
given by
$$n_{com}(z)
({\rm cm^{-3}})\;
\cong\;  {1.2\times10^6{\rm~s}\over 1+z}\times \dot n_{com}(z){\rm ~cm^{-3}~s^{-1}}\;,$$
with the comoving rate density, $\dot n_{com}(z)$, of blazar flares. 
This replacement should be accurate to a duty cycle factor of order unity.
Better studies based on GLAST observations will 
reveal the flaring behavior of $\gamma$-ray blazars; 
note that the threshold flux $f^{thr,G}_\epsilon(t)$  for
GLAST in the scanning mode is time-dependent when 
using eqs.\ (\ref{ddotNdz_1}) and (\ref{ddotNdz})
to fit the data.  

\section{Results}
\begin{figure}[b]
\epsscale{1.0}
\plotone{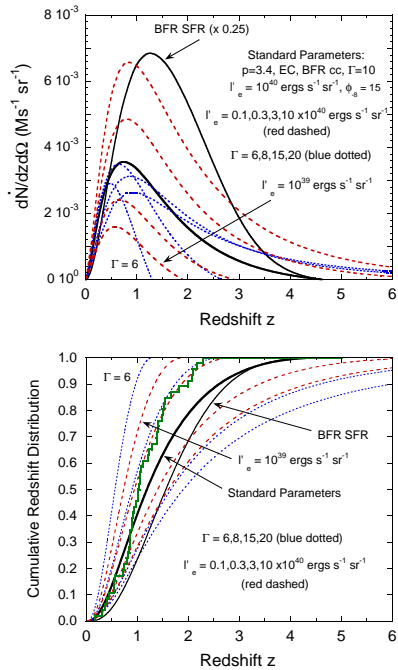}
\caption{Parameter study of the redshift distribution of FSRQs,
showing the effects of different parameter choices on the 
differential redshift distributions (top) and cumulative
redshift distributions (bottom).}
\label{fig4}
\end{figure}
Fig.\ 4 shows calculations of FSRQ blazar redshift distributions using the standard
FSRQ blazar parameter set, eq.\ (\ref{standardfsrq}), with differences
from the standard parameters as labeled. 
In this calculation, the detection threshold is 
$\phi_{-8} = 15$, corresponding
to a $\nu F_\nu$ flux threshold sensitivity, from eq.\ (\ref{fthr}), of 
$2.88\times 10^{-12}$ ergs cm$^{-2}$
s$^{-1}$ at $\epsilon = 200$ for $p = 3.4$. 
Except for the light solid curves, which show results
for a BFR given by the SFR history, eq.\ (\ref{SFR}), 
FSRQ blazars are assumed to emit blazar flares 
in random directions with a constant comoving rate density, 
The 
local rate density, $\dot n_{com}(z \ll 1)$, is set equal to $10^{-90}$ cm$^{-3}$
s$^{-1}$.  

The effect of increasing $\Gamma$ or $l^\prime_e$, of course, is to increase
the distance from which blazars can be detected. Fig.\ 4a shows that
 the directional event rate per unit redshift is not greatly increased
with increasing $\Gamma$ factor, whereas this rate is increased
with increasing $l^\prime_e$. The reason for this is that by increasing $\Gamma$,
the emission is jetted into a smaller solid angle, rendering a smaller 
fraction of the blazars visible, though from a larger distance. Nevertheless,
the cumulative redshift distributions for different combinations of $\Gamma$ and $l^\prime_e$
can be very similar, as seen in Fig.\ 4b. When $\Gamma \gg 1$, these two quantities enter into the 
threshold cosine angle $\hat\mu$ according to the combination ${l^\prime_e}^{1/q}/\Gamma$ (eq.\ [\ref{hatmu}]), 
reflecting the degeneracy of the results as a function of these two quantities. 

By changing the BFR function, the distribution of blazars with redshift can 
be adjusted to better agree with the observations. 
As is made clear by Fig.\ 4b, the effect of using the SFR function rather
than the constant comoving rate is to decrease the number of low-redshift blazars, 
and to have more blazars detected at $1 \lesssim z \lesssim 3$.

\begin{figure}[b]
\epsscale{1.0}
\plotone{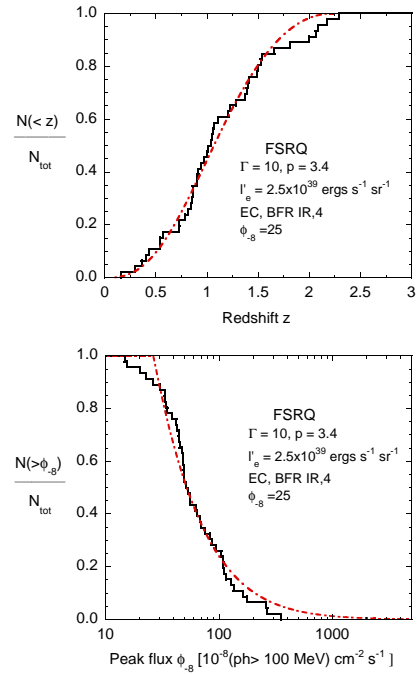}
\caption{Smooth curves give model fits to the cumulative redshift and 
size distributions of FSRQ EGRET data, shown by the histograms 
in the top and bottom panels, respectively. Parameters for
the fit are shown in the legend to the figure.}
\label{fig5}
\end{figure}

Even though it is relatively simple to find a parameter set
and BFR that gives a good fit to the FSRQ redshift distribution, it is 
considerably less simple to find a parameter set that gives
a good fit to the joint redshift
and size distribution. One difficulty in fitting the size 
distribution originates from sample
incompleteness near the EGRET threshold. The value $\phi_{-8} =15$
applies to two-week observations of on-axis EGRET sources (Appendix A). 
Reduction in the effective area for off-axis sources means
that the effective threshold where the EGRET sample is complete
is larger than $\phi_{-8} =15$. As a consequence, in the 
fitting of the data for FSRQs and BLs, the flux threshold was changed to $\phi_{-8} =25$ 
where the EGRET sample evidently no longer suffers from incompleteness. 
After adjusting the parameters and BFR histories, an acceptable 
fit to the FSRQ was obtained, as shown in Fig.\ 5.

This fit uses BFR IR,4 shown in Fig.\ 3, with parameter
values shown in the figure caption. The value of $p =3.4$ was chosen to 
to agree with the mean FSRQ photon spectral index $\alpha_{ph} = 2.2$ \citep{muk97}.
The values of $\Gamma$ and $l_e^\prime$ are not unique due to the degeneracy 
mentioned above, but the choice of $\Gamma = 10$ is suggested by superluminal 
radio observations of FSRQs \citep{up95}. As can be seen, the model fits the 
cumulative size distribution at $\phi_{-8} > 25$ but not at lower values
of $\phi_{-8}$ due to sample incompleteness.
The FSRQ model 
gives a statistically acceptable fit to the distributions, 
noting that the Kolmogorov-Smirnov one-sided statistic for 46 sources 
(42 sources with $\phi_{-8} = 25$.) is
0.16 and 0.22 at the 90\% and and 99\% confidence level, respectivly. 
The model fit, does, however, show a slight 
deficit of blazars at $1.5\lesssim z \lesssim 2$ compared to the data.
Moreover, the model somewhat overproduces the number of very bright FSRQs. 
Both the high-redshift deficit and overproduction of the brightest FSRQs
could be alleviated by tuning the BFR to increase even faster than given
by the model BFR, inasmuch as the brightest sources are generally found at lower
redshifts. Given the statistically acceptable fit, further fine-tuning
would, however, introduce additional parameters that are not well constrained. 

Even after searching over a wide range of BFRs for the BLs, it
was not found possible to obtain acceptable fits to the EGRET
data for the joint redshift and size distributions of BLs. The
difficulty was the very narrow range of peak fluxes, spanning
less than a factor $\approx 5$ from the dimmest to the brightest
values (compared to a factor $\approx 20$ for the FSRQs), and 
the requirement to have the same threshold integral photon
flux $\phi_{-8}=25$. The crux of the problem is that
the blazar size distribution is steeper 
than $-3/2$ (Fig.\ 2). A resolution of this problem was to introduce
luminosity evolution of the BLs such that they dimmed with 
increasing time. At the same time, the comoving rate density 
of BLs increases with time, allowing for a large number of nearby
BLs to be observed so that the redshift distribution of BLs could 
be fit. The effect of dimming luminosity  and 
increasing rate density with time makes 
blazar fluxes over a narrow range of values.

\begin{figure}[b]
\epsscale{1.0}
\plotone{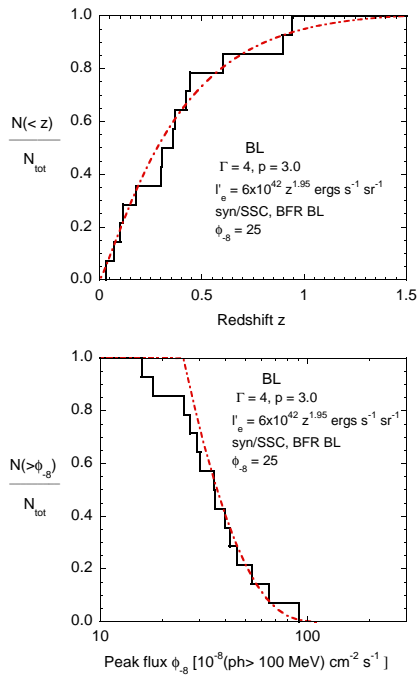}
\caption{Smooth curves give model fits to the cumulative redshift and 
size distributions of BL EGRET data, shown by the histograms 
in the top and bottom panels, respectively. Parameters for
the fit are shown in the legend to the figure. The rate-density 
dependence of the BL Lac flares is $\propto z^{-1.75}$.}
\label{fig6}
\end{figure}

Fig.\ 6 shows the result of this procedure, using eq.\ (\ref{nBLz}) to 
describe the BFR of BLs, with $a = 1.75$, and luminosity evolution
described by $l_e^\prime \propto z^{1.95}$. We choose $p = 2$ to agree
with the mean BL photon spectral index $\alpha_{ph} = 2.0$ \citep{muk97}, 
which implies a threshold $\nu F_\nu$ EGRET flux of $f_\e^{thr} = 4\times 
10^{-11}$ ergs cm$^{-2}$ s$^{-1}$ for $\phi_{=8} = 25$.
Other parameters of this
model are given in the caption to Fig.\ 6. The model 
gives a statistically acceptable fit, 
noting that in this case, the Kolmogorov-Smirnov one-sided statistic for 14 sources 
(12 sources for $\phi_{-8} = 25$.) is
0.275 and 0.39 at the 90\% and and 99\% confidence level, respectivly. 

The sample of FSRQs and BLs with $\phi_{-8} > 25$ 
suffers much less from sample incompleteness near threshold.
This flux-limited sample has 4 fewer FSRQ and 2 fewer BL sources,
and is therefore somewhat less constraining to the model fits. Although the fits 
to the size distribution using this smaller sample would
be improved near threshold, the modified redshift distribution is 
not significantly changed, and the use of this sample does not change 
the conclusions of this study.

\section{Predictions}

Fig.\ \ref{fig1} shows the fits to the FSRQ and BL redshift
distributions implied by the models 
discussed in the previous section. The
factors used to normalize to the flaring rates give a local FSRQ flare rate density equal to 
$$\dot n_{FSRQ} \cong 5.66\times 10^{-88} {\rm~cm}^{-3}{\rm~s}^{-1}\cong $$
\begin{equation}
 17 {\rm ~Gpc}^{-3}{\rm~Ms}^{-1}
\cong  1.7\times 10^{-8} {\rm ~Mpc}^{-3}{\rm~Ms}^{-1}\;
\label{dotnfsrq}
\end{equation}
for FSRQs, and a comoving BL flare rate density at $z = 1$ equal to
$$\dot n_{BL}(z = 1) \cong 1.23\times 10^{-88} {\rm~cm}^{-3}{\rm~s}^{-1}\cong $$
\begin{equation}
3.6 {\rm ~Gpc}^{-3}{\rm~Ms}^{-1}\;
\cong  3.6\times 10^{-9} {\rm ~Mpc}^{-3}{\rm~Ms}^{-1}\;
\label{dotnbl}
\end{equation}
for BLs.
The predicted size distributions of FSRQs and BLs extrapoloated
to small values of $\phi_{-8}$ implied by the model 
are shown by the solid curves in Fig.\ \ref{fig2}.  
Also shown is the one-year sensitivity of GLAST in the scanning
mode, which is at the level of $\approx 0.4\times 10^{-8}$ 
ph($>100$ MeV) cm$^{-2}$ s$^{-1}$. The model fit
predicts that GLAST will detect $\approx 700$ FSRQs and 
FR2 radio galaxies, and $\approx 160$ BLs and FR1 radio galaxies
after one year of observation, and an additional $\approx 200$ dim
FSRQs/FR2 galaxies and $\approx 50$ more BLs/FR1 galaxies after 3 more years of 
observation.
These numbers should be taken with an estimated uncertainty less than a factor-of-2, 
considering our lack of knowledge of blazar activity at
 $z \gtrsim 3$. The predicted numbers  would represent a lower
limit if there are blazar subclasses yet to be discovered that are numerous.
Broadening of the luminosity and $\Gamma$ distributions might increase the total
number, though by only factor of order unity.
Additional classes of low luminosity blazars that were not detected with 
EGRET could be discovered with GLAST, and these would produce an 
upturn in the size distribution at low fluxes. The number of such 
sources are, of course, limited by the level of the EGRB.

\begin{figure}[b]
\epsscale{1.0}
\plotone{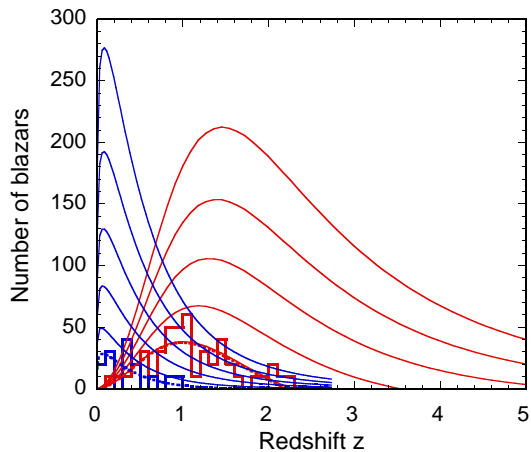}
\caption{Histograms show the redshift distributions of FSRQs and 
BLs discovered with EGRET. Dotted curves show the model fits to the
distributions, using detection characteristics typical of EGRET
(and $\phi_{-8} = 25$; see Fig.\ 1). Solid curves show the predicted FSRQ 
and BL redshift distributions using the parameters used to model
the EGRET data, though with detection characteristics
of GLAST. For the GLAST predictions, the observing photon energy
 is set equal to 1 GeV ($\e \approx 2000$), and the detection 
sensitivity is 1, 3, 10, 30, and $100\times$ better than EGRET, 
using the $\phi_{-8}= 15$ as the flux threshold for EGRET. }
\label{fig7}
\end{figure}

Fig.\ 7 shows the predicted FSRQ and
BL blazar redshift distributions for GLAST.  The distribution
predicted for EGRET is shown by the dashed curves, corresponding
to the result shown in Fig.\ 1,
and the predicted distributions for GLAST sensitivities of $ 1, 3,
10, 30,$ and  $100\times$ the EGRET sensitivity of $\phi_{-8} = 15$ are
shown by the solid curves. 

\begin{figure}[t]
\epsscale{1.3}
\vskip-1.8in
\plotone{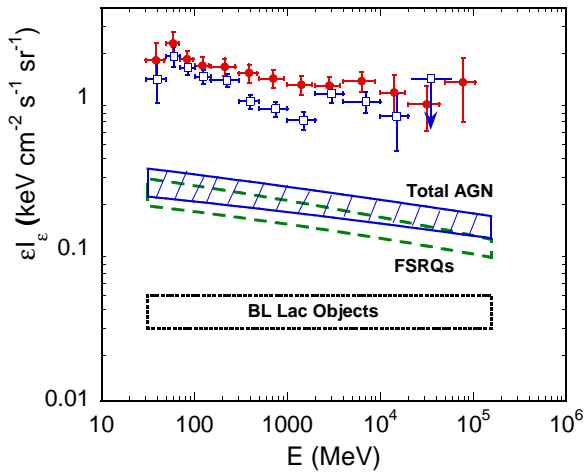}
\caption{The intensity of the isotropic EGRB measured with 
EGRET is shown by the filled data points \citep{sre98}, and the
EGRB after subtraction of a component of quasi-isotropic galactic $\gamma$ 
radiation is shown by
the open data points \citep{smr04}. The dashed box gives the diffuse
intensity from unresolved FSRQs, and the dotted box give the
diffuse intensity from unresolved BLs, and the curve labeled 
``Total AGN" is the sum of the two.  The boundary of the box represents the flux
limit below which blazars are assumed to be unresolved, and correspond
to the range for $\phi_{-8} = 25$ and $\phi_{-8} = 12.5$.}
\label{fig8}
\end{figure}

Fig.\ 8 shows the contribution to the EGRB as reported by the
EGRET team \citep[filled data points;][]{sre98}, and as derived from
the EGRET data using the GALPROP cosmic-ray propagation code
\citep[open data points;][]{smr00,smr04}.  The boxes give 
the contributions to the EGRB from unresolved FSRQs
and BLs with flux values between $\phi_{-8} = 25$, below which
a two-week observation is incomplete, and $\phi_{-8} = 12.5$. 
This latter value is a factor of two less than the brighter value,
and is chosen because EGRET had an effective lifetime of $\approx 4$
years in terms of its capability during the first year, due to 
limited amount of spark chamber gas. This would produce a factor $\approx
2$ better sensitivity before sample incompleteness dominates. 
Nonuniformity in exposure and angle to the normal of EGRET
suggests that the threshold is in fact quite broad.

The steeper FSRQ and flatter BL spectra
in Fig. 8 arise from using $p = 3.4$ for 
FSRQs and $p=3$ for BL Lacs (note also 
the weak curvature in the calculated FSRQ background
spectrum). 
The model results imply that the 
total AGN (BL$+$FSRQ) contribution to the EGRB as measured by \cite{sre98} 
is at the level of $\approx 12$ -- 20\%, and has a spectral index
remarkably similar to the index measured in the EGRET analysis. The FSRQ
contribution to the EGRET  $\gamma$-ray background at 1 GeV is
estimated at the level of $\approx 10$ -- 15\%, 
and the BL contribution is $\approx 2$\% -- 4\%. This contribution also 
includes FR1 radio galaxy, estimated by \citet{sta06} as contributing
at the $\approx 1$\% level ($\approx 6\times 10^{-3}$ keV cm$^{-2}$ s$^{-1}$ sr$^{-1}$.
The other sources classes that contribute to the residual EGRB are 
considered elsewhere \citep{der06}. Note that no hard component
BL Lac contribution to the EGRB is calculated here, though 
one must exist given measurements of hard-GeV spectrum 
BL Lac objects \cite[e.g.,][for Mrk 501, which is not in our sample]{kat99,pet00}. Spectral hardening during 
flaring is also not considered here. Even so, this calculation 
implies that there is room for a number of other sources and/or 
source classes to make up the EGRB.

\section{Discussion}

We have employed a simple physical model to fit population statistics of
EGRET $\gamma$-ray blazar data and make predictions for the data anticipated from GLAST.  
By characterizing all FSRQs and BLs in terms of single values
of $\Gamma$ and $l^\prime_e$, the model contains the 
fewest number of free parameters while still agreeing with the 
basic physical understanding of blazars as a two-sided jet of 
collimated relativistic plasma outflow with the radiation
Doppler boosted by the  motion of the jet.  Other than blazar 
rate density evolution, described in terms of the BFR
 function, only the parameters $\Gamma$, the proper
directional power $l^\prime_e$, the power law index $p$ and 
beaming statistic $q$ define a model. The values of $p$ and $q$ 
are known with some certainty from observations. The model 
is further simplified in that the cumulative distributions
are degenerate in the quantity ${l^\prime_e}^{1/q}/\Gamma$
when $\Gamma \gg 1$.

We have used $\Gamma \cong 10$ for FSRQs and $\Gamma \cong 
4$ for BLs. These
values of $\Gamma$ are generally consistent with Lorentz factors
inferred from fits to the local luminosity functions of radio 
galaxies and blazars 
\citep{up95} and measurements of superluminal motions
\citep{vc94}. The superluminal motion observations indicate 
that BLs have smaller Doppler factors on average than FSRQs, though 
in both cases the spread in Doppler factors is rather large \citep{jor01}.
Good fits to the FSRQ data are obtained
with parameters $\Gamma = 10$, $l_e^\prime = 2.5\times  10^{39}$ 
ergs s$^{-1}$ sr$^{-1}$, $p = 3.4$, and the EC beaming statistic. 
A BFR function that increases by a factor $10$ -- 20 
from $z \ll 1$ to $z\approx 1$ -- 2
was required to jointly fit the FSRQ redshift and size distribution, but
a good fit was obtained even without luminosity evolution. 

In contrast, it was not found possible to obtain a good fit to the 
BL redshift and size distributions without also invoking luminosity evolution.
The BL size distribution declines more rapidly than a uniform Euclidean
size distribution $N(>\phi)\propto \phi^{-3/2}$, as can be 
seen in Fig.\ 2. Any luminosity function, whether a kinematic 
luminosity function produced by the randomly 
oriented blazar jets, or a luminosity function given in terms of 
a range of jet powers, only flattens
the size distribution. A feasible solution to this dilemma was to have
the BL luminosity dim with time while the rate density of BL flares increases with time.
A model that provides a reasonable fit to the BL data
has $\Gamma = 4$, $l_e^\prime (z) = 6\times 10^{42}z^{1.95}$
ergs s$^{-1}$ sr$^{-1}$, a rate density evolution of BLs $\propto z^{-9/4}$, 
$p = 3.0$, and the synchrotron/SSC beaming statistic.  

Before proceeding to the implications of this analysis, a few
points should be made.
\begin{enumerate}
\item The tightly constrained parameter space in this simplified blazar model 
leads to fairly robust predictions for the number and intensity of blazars,
and we think that the number of blazars and radio galaxies cannot be more than 
a factor-of-two in error, and then in the direction of more sources and
greater AGN background $\gamma$ rays. A more precise error analysis must be
considered in further
study of this model.
\item Luminosity evolution was introduced in the BL fitting, but not the 
FSRQ fitting. The blazar's power 
evolves with cosmic times, reflecting the growth of the supermassive
black hole (including binary black-hole merger episodes) and its intermittent fueling.
The jet power is related to the accretion power in ways not fully
understood, and is likely also dependent on black-hole spin. Modeling FSRQ growth
and fueling with only a BFR function is possible with the EGRET data, but
may well fail when modeling the larger and more meaningful GLAST data set. 
The next step in the analysis effort is to allow luminosity evolution of both
FSRQs and BLs based on a physical model for supermassive black-hole growth
and fueling, including duty factors on small and large timescales.
\item The quality of the fits, though acceptable, is not as good
as in the study of \citet{mp00}. This indicates that a range of luminosities
or injection energies may be required to obtain the best fits for the FSRQs, 
and that the form of the 
$\gamma$-ray luminosity function may  be
well represented by the \ luminosity function  of FR1 and 
FR2 radio galaxies.
Note also that the BL sample used in 
the \citet{mp00} appears smaller and less constraining than the BL sample
used here. The size distributions of 
BLs measured with 
GLAST will show whether the 
BL sample used in this study is representative. 

\end{enumerate}

\subsection{Blazar evolution}

The solution presented here that fits both the steep BL size distribution and 
the BL redshift 
distribution was to have joint {\it positive} luminosity evolution
and {\it negative} source density evolution. From a simple size-distribution
analysis in the Euclidean limit, the flux ($\phi$) dependence of the 
integral size distribution goes as   
$$N(>\phi) \propto \phi^{-(3-a_n)/(2-a_l)}\;,$$
where $l_e^\prime \propto z^{a_l}$ and $n(z)\propto z^{-a_n}$, and $a_n= 0$ for a uniform
distribution of sources. A flatter (steeper) size distribution occurs for negative (positive) luminosity 
evolution with $a_l < (>) 0$. Arranging to fit not only the BL size distribution with slope steeper than 
the uniform Euclidean value of $-3/2$ requires positive luminosity evolution, and also to
fit  the measured BL
redshift distribution required negative source rate-density evolution, leading after 
trial and error to the result
 $a_l = 1.95$ and $a_n = 9/4$. Recalling the 
definition of $\langle V/V_{max} \rangle$ as the normalized sum over the test statistic 
 $(\phi_i/\phi_{thr})^{-3/2}$, one sees that size distributions
steeper than $-3/2$ have a larger number of faint detections near threshold than 
for a uniform Euclidean distributions and thus have $\langle V/V_{max} \rangle > 0.5$;
size distributions flatter than $-3/2$ have a deficit of faint detections near threshold
 and thus have $\langle V/V_{max} \rangle < 0.5$. 

This EGRET BL analysis indicates that $\langle V/V_{max} \rangle > 0.5$, 
similar to the case for bright 1 Jy radio sources or the brightest BL Lac 
objects selected at X-ray energies \citep{gio01}. From Fig. 2, we predict that 
 $\langle V/V_{max} \rangle $ for BLs will shift to $<0.5$ within a few months
of the start of GLAST measurements, as fainter BL Lac objects and FR1 radio galaxies
will be lost due to cosmological effects and the small number of high-redshift
BLs.  Our analysis leads, however, to very different conclusions
than obtained previously when considering only the size distribution and $\langle V/V_{max} \rangle$ 
value from 
various BL surveys at radio and X-ray energies. 
For example, a flatter than $-3/2$ size distribution 
is found in the ROSAT (0.5 -- 2.0 keV) all-sky survey (RASS) of over 30 BL Lac objects
complete to $f_x(0.5$ -- 2.0 keV$) > 8\times 10^{-13}$ ergs cm$^{-2}$ s$^{-1}$ \citep{bad98},
corresponding to
a deficit of sources near threshold, or negative BL luminosity evolution. 
The same conclusion is found by \citet{gio99}, who assemble a  sample of 155 BL Lacs
by cross correlating the RASS survey results with radio sources from the NRAO VLA Sky Survey.
The resulting radio luminosity size distribution of these X-ray selected BL Lac objects
also exhibit a negative 
BL evolution.  \citet{rec00} measured redshifts and analyzed sources
in the Einstein 0.3 -- 3.5 Medium Sensitivity Survey, and obtained
 $\langle V/V_{max} \rangle \cong 0.4,$
again showing a deficit of near threshold sources compared to the uniform, Euclidean 
expectation. 

It is meaningful to interpret a flattening of the size distribution compared to a $-3/2$ slope
as a reduction in luminosity at early times \citep{pg95a}. 
Here however we additionally require, as do \citet{mp00},
that the redshift distribution is also accurately fit with the same model that fits the size
distribution. Unlike the  \citet{gio06}
calculation of the EGRB from 
$\gamma$-ray blazars, no radio/$\gamma$-ray connection is assumed throughout this
analysis (though radio data is used to guide the choices of $\Gamma$). 
The results from joint fitting of EGRET data are severely
constraining: the comoving rate density of FSRQ declines by a factor $\gtrsim 10$ 
at  $z \ll 1$ compared to $z \lesssim 1$ -- 2, while
 the rate density of BLs rapidly increases for $z \ll 1$ and, at the same
time, the mean 
luminosity of BL flares declines with time. 

This behavior is in striking accord with the scenario \citep{bd02,ce02}
linking BLs to FSRQs through an evolution in declining accretion
rates and increasing black hole masses. In the particular formulation of \citet{bd02},
the reduction in the amount of gas and dust that both fuels the central 
black hole engine and scatters accretion-disk radiation causes
a transformation of the FSRQs into the BLs in terms of spectral properties, 
thereby explaining the blazar
sequence correlating $\nu F_\nu$ peak synchrotron frequencies and apparent synchrotron 
power \citep{sam96,fos98}, and puts on a more physical basis
the correlations between the peak synchrotron frequency, peak electron 
Lorentz factor, and injection and external radiation compactnesses proposed
by \citet{ghi98} to explain the blazar sequence.   \citet{ce02} argue that the mean 
luminosity of the BLs does not change much over several gigayears,
and that the flattened size distribution of BLs at low fluxes
stem from negative density evolution, with spectral differences
associated with transition from accretion power to a component from
the black-hole spin. 

A clearcut prediction of this scenario is that the mean masses of
black holes in BLs is larger than that in FSRQs at the same epoch.
The minimum variability timescale of blazar flares should be proportional
to black hole mass (\S A.5), but the shorter timescale
flaring behavior of BL Lacs measured with Whipple and HESS compared
with the minimum variability 
timescale of FSRQs measured with EGRET is opposite to the expected behavior, 
but may only reflect sensitivity limitations of EGRET.  Probably there
is flaring on every timescale, and a power spectral density analysis
is required to see if there is a size scale where temporal power 
declines, corresponding to the black hole mass. 
This question will obviously be subject to extensive
investigation with GLAST.  

\citet{gc01} infer the black hole masses of FR1 and FR2 radio 
galaxies on the basis of an expression for the central 
black hole mass in terms of host galaxy  absolute
R-band magnitudes. Relating jet power to radio luminosity
suggests that FR1 galaxies are low Eddington-ratio ($\lesssim 10^{-2}$)
and FR2 galaxies are moderate Eddington-ratio  ($0.01$ -- $0.1$) sources.
There is a tendency for the FR1 galaxies to have larger black-hole masses than 
FR2 galaxies, but this effect is too weak to provide an unambiguous test of the 
$FSRQ\rightarrow BL$ evolutionary 
scenario.

It is unphysical to extend the form of 
the BL rate density evolution, $\propto z^{-9/4}$,
to arbitrarily small redshifts. The comoving
 density of  FSRQs forming at high 
redshifts should equal the comoving density of the BLs 
at low redshifts in the picture of FSRQ $\rightarrow$ BL evolution
considered here. The BL density will saturate
at $z_{bl}$
when all FSRQs have converted into BLs. If the duty cycle of the 
two sources classes is similar, this occurs at 
$\dot n_{BL}(z_{bl}) \cong \dot n_{FSRQ} (z \gtrsim 1)$. Using
the derived values in eqs.\ (\ref{dotnfsrq}) and (\ref{dotnbl}) and the 
BFR functions, this takes place at $z_{bl} \cong 0.1$. Modifying
the BL BFR to be constant at $z \leq z_{bl}$ has minimal effects on the 
results, but should be considered in the next step of the analysis.

\subsection{Space density of blazars}

The local density of blazar sources given from this analysis
can be compared with the space density of blazar host galaxies, 
namely early-type elliptical galaxies.  From a K-band survey of 
bright galaxies with $z \lesssim 0.4$, \citet{hua03} calculate
a local galaxy  density of $0.0048\pm 0.001$ Mpc$^{-3}$ for 
galaxies of all types with $h = 0.72$. The density of early-type galaxies is
$\approx 10$\% of the total, or  $\approx 5\times 10^{-4}$ Mpc$^{-3}$. 
In comparison, the comoving densities of the sources of FSRQ and BL flares 
are $n_{FSRQ}(z\ll 1) \cong 2\times 10^{-8}$ Mpc$^{-3}$ and 
$n_{BL}(z = 1) \cong 2\times 10^{-9}$ Mpc$^{-3}$, assuming a duty
factor equal to unity (a duty factor less than unity
implies a proportionately smaller source density).

The comoving density of sources at the present epoch is, in the 
scenario where FSRQs evolve into BLs, equal to $\approx \Sigma_{FSRQ}(z\gg 1)n_{FSRQ} (z \ll 1)
\cong 17 n_{FSRQ} (z \ll 1)\cong 4\times 10^{-7}$ Mpc$^{-3}$.
This is about three orders of magnitude smaller than the elliptical
galaxy density. Either only 
a small fraction of ellipticals host radio galaxies, or the duty 
cycle of elliptical galaxies that harbor radio and $\gamma$-ray
jets is a small fraction of the total lifetime of the source \cite[see][for 
more speculations]{hco04}.

\section{Summary and Conclusions}

A method to analyze the population 
statistics of $\gamma$-ray blazars solely from the $\gamma$-ray 
data was considered in this paper \citep[see also][]{mp00}. 
By performing the $\gamma$-ray population study
independent of other wavebands, the reliability of this method to
other methods that invoke a radio/$\gamma$-ray correlation can be tested, and 
the physical reasons for differences
in radio galaxy and blazar populations selected from observations at different 
wavebands can be explored. 

Although a radio/$\gamma$-ray correlation was avoided,
guidance to assign values of $\Gamma$ was taken from radio 
observations. Measurements of apparent superluminal motion 
from radio observations on
the scale of $\approx 0.1$ -- 1 pc from the black hole 
imply $\Gamma$ factors. The $\gamma$
rays may originate from within hundreds to thousands of Schwarzschild
radii of the supermassive black hole, so that the $\Gamma$ values
derived from radio observations may not be appropriate to the analysis
of $\gamma$-ray data. Radio observations suggest that
$\gamma$-ray flares could originate from the same physical scale as the
superluminal blobs \citep{jor01a}, though
spectral modeling of some X-ray selected BLs such as
Mrk 421 and Mrk 501 suggest larger Doppler factors $\approx 50$
\citep[][but see \citet{gk03,gtc05} for possible 
resolutions to this puzzle]{kra01} than inferred from
superluminal motion observations. 
Lower limits to the values of $\Gamma$ for BLs and FSRQs will be
inferred from $\gamma\gamma$ attenuation arguments applied to the
GLAST observations.  This will not only provide better values to use
for modeling the population statistics of $\gamma$-ray blazars, but
will help break the parameter degeneracy in the modeling.

By avoiding an underlying radio/$\gamma$-ray assumption,
only a very simplified blazar model could be investigated.
This model nevertheless 
contains the essential blazar physics, and can be simply generalized
to include a range
of $\Gamma$ factors and an evolving, broadened luminosity function. 
For the analysis of the EGRET data, introducing distributions
in $\Gamma$ or $l_e^\prime$ would allow too large a parameter space to explore
without introducing some radio/$\gamma$-ray connections, so 
we considered fixed values of $\Gamma = 10$ for FSRQs and $\Gamma = 4$ for BLs, and
found values of luminosity $l_e^\prime $ that gave reasonable fits to the data.
Because mean spectral indices were assigned from $\gamma$-ray observations, and 
the beaming factor statistic was taken from blazar physics, only the value 
of $l_e^\prime $, the BFR function and, when necessary,
the redshift dependence of $l_e^\prime$,
 were varied in order to fit the EGRET redshift and size distributions. 

Within these constraints, the EGRET data for FSRQs was fit with a BFR function
that had $\approx 10\times$ more sources at $z = 1$ than at present, 
which could be related either to a star-formation rate function
or an evolutionary behavior proportional to the far
IR/sub-millimeter luminosity density related to IR radiation
\citep{san04,bla00}. The EGRET
data for BLs could not be fit by only modifying the form 
of the BFR, and luminosity evolution was also required. The resulting 
solution---that the density of BLs increase and  their mean jet powers dim
with time---is in accord with the scenario where FSRQs evolve from BLs.

The analysis implies that the contribution 
of unresolved blazars to the extragalactic $\gamma$-ray background measured
with EGRET \citep{sre98}
at the level of $\approx 20$\% at 1 GeV, leaving room for various source classes
or additional types of blazars, such as hard GeV-spectrum blazars, to which EGRET
was not very sensitive. Approximately 1000 blazars are predicted from this analysis to be 
discovered with GLAST during its first year of operation, but this could be an underestimate
for the reason just stated, though not by more than a factor of 2. As also observed with EGRET, the 
BL/FR1 $\gamma$-ray sources are predicted to have a much smaller mean redshift than
the FSRQ/FR2 sources (Fig.\ 7). The predictions for the 
high redshift ($z \gtrsim 3$ -- 5) blazar population is very uncertain because
of the very few high-$z$ $\gamma$-ray blazars observed with EGRET.
If blazars only make $\simeq 20$\% of the EGRB at 1 GeV, then star-forming galaxies,
 starburst galaxies, and cluster of galaxies are likely to contribute a significant
fraction of the difference, so we can expect GLAST to discover new $\gamma$-ray 
source classes. 

With the much larger data set from GLAST, 
more detailed analyses will allow various effects related to blazar 
spectra and flaring and the evolutionary connections between various
classes of cosmological black-hole jet sources to be examined in much greater detail.

\acknowledgments
I thank Stanley P.\ Davis for helping assemble the sample used in the
analysis. I would also like to thank
J.\ Chiang, S.\ Ciprini, B.\ Dingus, G.\ Fossati, P.\ Giommi, T.\ Le, G.\
Madejski, J.\ McEnery, F.\ C.\ Michel,   and R.\ Romani for useful comments and
discussions that influenced this work. Thanks are also due to B.\ Lott, 
P.\ Michelson, and S.\ Ritz for encouragement, to P.\ Sreekumar for
providing the EGRET data for the diffuse $\gamma$-ray background, and particular thanks
to A.\ Reimer and the anonymous referee for many helpful comments and useful suggestions.  
The work of C.\ D.\ D.\ is supported by the Office of Naval Research and
GLAST Interdisciplinary Science Investigator grant DPR S-13756G.

\appendix

\section{Sensitivity of EGRET and GLAST to Blazar Flares}

We follow the approach of \citet{dd03}.  The properties of a
$\gamma$-ray imaging spark chamber or silicon tracker detector depends
on its shower pattern, assumed to be described by a Gaussian with 68\%
containment angle $\hat\theta$ and energy-dependent angular point
spread function (psf) described by
\begin{equation}
\theta_{psf} = \hat \theta u^{-w}\;,
\label{thetapsf}
\end{equation}
where the photon energy in units of 100 MeV is
\begin{equation}
u \;=\;{E\over E_{100}}\;\;{\rm and}\;\;E_{100} = 100\;{\rm MeV}\;.
\label{eovere100}
\end{equation}
For EGRET, $\hat\theta \cong 5.7^\circ/57.3^\circ \cong 0.1$ and $w =
1/2$.  For GLAST, $\hat\theta \cong 3.5^\circ/57.3^\circ \cong 0.06$
and $w = 2/3$.

The effective area $A(E,\theta,\phi)$ depends on photon energy $E$,
angle $\theta$ from zenith, and azimuthal angle $\phi $ measured with
respect to the zenith angle, according to
\begin{equation}
A(E,\theta,\phi) \;\cong \; A(u) \cong A_0 u^a_0\;,
\label{Au}
\end{equation}
where $A_0$ is the effective area averaged over the field of view,
defined as the opening solid angle within which the effective area is
50\% of the on-axis FOV.  The FOV of EGRET is $\cong 4\pi/24 \cong
0.5$ sr, and the FOV of GLAST is $\cong 4\pi/6 \cong 2$ sr.  The
parameters describing the effective area $A_0$ at 100 MeV and index
$a_0$ are, respectively, $\cong 1200$ cm$^2$ and $a_0\cong 0$ for
EGRET, and $\cong 6200 $ cm$^2$ and $a_0 \cong 0.16$ for
GLAST.\footnote{These satisfy the GLAST Science Requirements Document;
the actual performance is given at the GLAST websites. Note that the
estimate requires an average over the FOV.}

The significance to detect a signal at the $n\sigma$ level is given by
\begin{equation}
n \cong {S\over \sqrt{\alpha S + (1+\alpha)B}}
\label{n}
\end{equation}
\citep[][eq.\ 9]{lm83}, where $S$ is the number of source counts, $B$
is the number of background counts, and $ \alpha = t_{on}/t_{off}$ is
the ratio of on-source to off-source observing times. If the
background in precisely known, $\alpha \rightarrow 0$, and
\begin{equation}
n \cong {S\over \sqrt{B}}\;.
\label{napprox}
\end{equation}
In the limit $\alpha \ll 1$ and $n \gg 1$, eqs.\ (1) and (2)
overestimate the fluctuation probability compared to a maximum
likelihood expression confirmed by Monte Carlo simulations
\citep{lm83}, so eq.\ (\ref{napprox}) should be a conservative
expression for the detection significance.
 
\subsection{Background Counts}

We assume that the source location is precisely known. The number of
background photons within solid element $\Delta \Omega(E)$ centered in
the direction $\vec\Omega$, and with energies $>E_1$ observed during
an observing time $\Delta t$ is
\begin{equation}
B(>E_1) \cong \int_0^{\Delta t}dt\int_{E_1}^\infty dE \;\Delta \Omega(E) \;
 A[E,\theta(t),\phi(t)]\Phi_B(E,\vec\Omega)\; .
\label{B}
\end{equation}
The background photon flux per steradian, $\Phi_B(E,\vec\Omega)$ $=
dN_\gamma /dAdtd\Omega dE$, is assumed to be time-independent; thus
this treatment excludes passage through the South Atlantic Anomaly.
This expression also applies to time-independent diffuse backgrounds,
not time-variable point sources or variable backgrounds.
The effective area $A[E,\theta(t),\phi(t)]$ of the 
telescope at energy $E$ changes for a source at time-varying angles $\theta$ 
and $\phi$ with respect to the telescope $\hat z$-axis,
which change with time due to rocking or slewing or Earth occultation. 
These effects are taken into account with an exposure factor $X$.
Note that eq.\ (\ref{B}) assumes that each photon count 
can be precisely assigned an energy E and direction $(\theta,\phi)$. 

The apparently diffuse isotropic $\gamma$-ray background spectrum 
is independent of $\vec\Omega$ and is given by 
\citet{sre98} as 
\begin{equation}
\Phi_X(E) = k_x u^{-\alpha_B}\;, 
\label{Phi_B}
\end{equation}
where $k_x = 1.73\pm 0.08 \times 10^{-7}$ ph (cm$^2$-s-sr-MeV)$^{-1}$
and $\alpha_B = 2.10\pm 0.03$. This expression is a valid description
of the minimum $\gamma$-ray background irregardless of whether the it is
extragalactic, or contains a quasi-isotropic 
galactic or heliospheric emission component
\citep{smr04}.  
Thus the number of background counts
with energy $E$ observed over the real time $\Delta t = t_{wk}$ weeks
is
\begin{equation}
B(>u) = {X\Delta t E_{100} \pi \hat\theta^2 A_0 k_x\over 
2w+\alpha_B -1 -a_0}\; u^{1+a_0 -2w -\alpha_B}\;,
\label{bgte}
\end{equation}

\subsection{Source Counts}

For a point source, the number of source counts with energy $>E$ is
given by
\begin{equation}
S(>E_1) \cong f_\gamma \int _0^{\Delta t} dt \int_{E_1}^\infty dE 
\; A[E,\theta(t),\phi(t)]\; \phi_s(E,t)\;,
\label{S}
\end{equation}
where $\phi_s(E,t)$ is the source photon flux (ph cm$^{-2}$ s$^{-1}$
E$^{-1}$) and we assume that the point spread function is defined as
the $f_\gamma = 68$\% containment radius
\citep[see][]{tho86}. 
Thus
\begin{equation}
S(>u) \cong {10^{-8} f_\gamma X_s \Delta t (\alpha_{ph} - 1) A_0 \phi_{-8}
\over
\alpha_{ph} - a_0 - 1} \;u^{1+a_0 -\alpha_{ph}}\;,
\label{se1}
\end{equation}
and $\alpha_{ph}$ is the photon number index (commonly denoted
$\Gamma$).  The factor $X_s$ accounts for the on-source exposure, and
is generally equal to the background exposure factor $X$, though it
may differ if one allows $X_s$ to account for the source duty cycle,
of if $X$ accounts for variable background. The quantity $\phi_{-8}$
normalizes the source flux
\begin{equation}
\phi_s(E) = {10^{-8} (\alpha_{ph} -1)\phi_{-8}\over
E_{100}}\;u^{-\alpha_{ph}}
\label{phise}
\end{equation}
to $10^{-8}$ ph($E> 100$ MeV) cm$^{-2}$ s$^{-1}$, which is the unit
quoted by the EGRET team in the Third EGRET catalog \citep{hea99} for
the two-week average fluxes.

For EGRET, $S(>u) \cong 4.9 u^{-1.1}\phi_{-8} t_{wk} \geq 5 n_{5s}$,
where a source detection is assumed to require at least $5$ counts
($n_{5s} = 1$).  From the signal limit, a minimum time
\begin{equation}
t^E_{wk}\gtrsim {u^{\alpha_{ph}-1}\over \phi_{-8} X_s}
\;n_5\;\cong \;{2u^{1.1}\over \phi_{-8}}\;,
\label{phim8E}
\end{equation}
is needed for EGRET to detect sources in the EGRET FOV as faint as
$\phi_{-8}$ (assuming that the blazar remains at the level of
$\phi_{-8}$ during the entire viewing period).  Earth occultation
means $X_s\approx 1/2$.  The final expression in eq.\ (\ref{phim8E})
applies when $\alpha_{ph} = 2.1$, the index of the isotropic
$\gamma$-ray background.  Here we suppose that blazars make a large
fraction of the EGRB, and so must have a significant fraction of
bright blazar sources emitting with $\alpha_{ph} \cong 2.3$ between
$\approx 100$ MeV and $10$ GeV.

For GLAST in the scanning mode, the condition for the detection of 5
counts is
\begin{equation}
 t^G_{wk}\gtrsim \left( {\alpha_{ph} - 1.16\over \alpha_{ph} - 1}\right)\;
{u^{\alpha_{ph} - 1.16}\over X_{1/5}\phi_{-8}}\; n_{5s} \;\cong\;
0.85\;{u^{0.94}\over  \phi_{-8}}\;,
\label{phim8G}
\end{equation}
where we have adopted a source occultation factor $X_s = 0.2 X_{1/5}$.
The two estimates are similar at $100$ MeV, because the factor 6 -- 8
effective area advantage of GLAST over EGRET is reduced by GLAST in
the scanning mode compared with the few occasions when the source is a
pointing target for EGRET.

\subsection{Signal to Background}

From eqs.\ (\ref{napprox}), (\ref{bgte}) and (\ref{se1}), the
significance to detect a blazar at the $n\sigma$ level is
\begin{equation}
n(>u) \; =\; {S(>u)\over \sqrt {B(>u)}}\; =\; 
{10^{-8} f_\gamma X_s \sqrt{A_0 \Delta t}  \phi_{-8}
\over
\sqrt{X\pi k_x E_{100}}\;\hat\theta}\;
\varphi_c \;u^{\varphi_x}\;,
\label{subu}
\end{equation}
where
\begin{equation}
\varphi_c = {(\alpha_{ph} - 1)\over
\alpha_{ph} - a_0 - 1}\;\sqrt{2w+\alpha_B -1-a_0}\;,
\label{varphic}
\end{equation}
and
\begin{equation}
\varphi_x = w - \alpha_{ph} + {1+a_0 +\alpha_B\over 2}\;.
\label{varphix}
\end{equation}

The energy dependence of the background-limited detections 
of blazars is defined by the index 
\begin{equation}
\varphi_x \cong \cases{2.05-\alpha_{ph},\; & EGRET
 $~$ \cr\cr 2.30-\alpha_{ph}\;, &GLAST$~$ \cr}\;,\; 
\label{varphixEG}
\end{equation}
and it is of interest that EGRET and GLAST are more sensitive to
hard-spectrum blazars ($\alpha_{ph}\lesssim 2$ -- 2.3) at higher
photon energies.  This sensitivity improves until either there is a
spectral break or softening, or the signal runs out.

The background limit on EGRET, with $X = X_s$, is
\begin{equation}
n^E(>u) \; =\; 0.36 \sqrt{X t_{wk}} \phi_{-8} 
u^{2.05 -\alpha_{ph}} \;>\; 5 n_{5\sigma}\;.
\label{neu}
\end{equation}
Taking $X = 0.5$, due to Earth occultation, the number of weeks of
direct pointing required for EGRET to detect a source with flux at the
level $\phi_{-8}$ at the 5$n_{5\sigma} \sigma$ significance level is
\begin{equation}
t^E_{wk} \gtrsim {386\over \phi_{-8}^2 } \; 
u^{2(\alpha_{ph} - 2.05)}\;n_{5\sigma}^2\;.
\label{twke}
\end{equation}
Comparison with eq.\ (\ref{phim8E}) shows, as is well known, that
signal detection with EGRET is background rather than signal limited.
For $u = 1$, as used in the analysis reported in the EGRET catalogs,
the limiting sensitivity to detect a strong source in a two-week
pointing is $$\phi_{-8} \gtrsim 14\;.$$ This estimate agrees well with
detection data, as seen from Fig.\ 2.

For GLAST, the background limit translates into the condition
\begin{equation}
\phi_{-8} \gtrsim {3.6 n_{5\sigma}\over 
\sqrt{Xt_{wk}}}\; u^{\alpha_{ph} - 2.3}\;,
\label{phig}
\end{equation}
so that the number of weeks of scanning
required for GLAST to detect a source with flux level $\phi_{-8}$ at the
5$n_{5\sigma} \sigma$ significance level is
\begin{equation}
t^G_{wk} \gtrsim {64 u^{2(\alpha_{ph} - 2.3)}  n_{5\sigma}^2 \over
 \; X_{1/5} \phi_{-8}^2}\;
\rightarrow 
{64\over \phi_{-8}^2 }\;u^{-2/5}\;.
\label{twkg}
\end{equation}
Thus, in around a year in the scanning mode, GLAST will detect sources
at the level $\phi_{-8} \sim 1$ when integrating above 100 MeV ($u =
1)$. This is not however the best detection strategy.

\subsection{Optimal Source Detection with GLAST}

\begin{figure}[t]
\vskip-1.0in
\epsscale{1.0}
\vskip-2.0in
\plotone{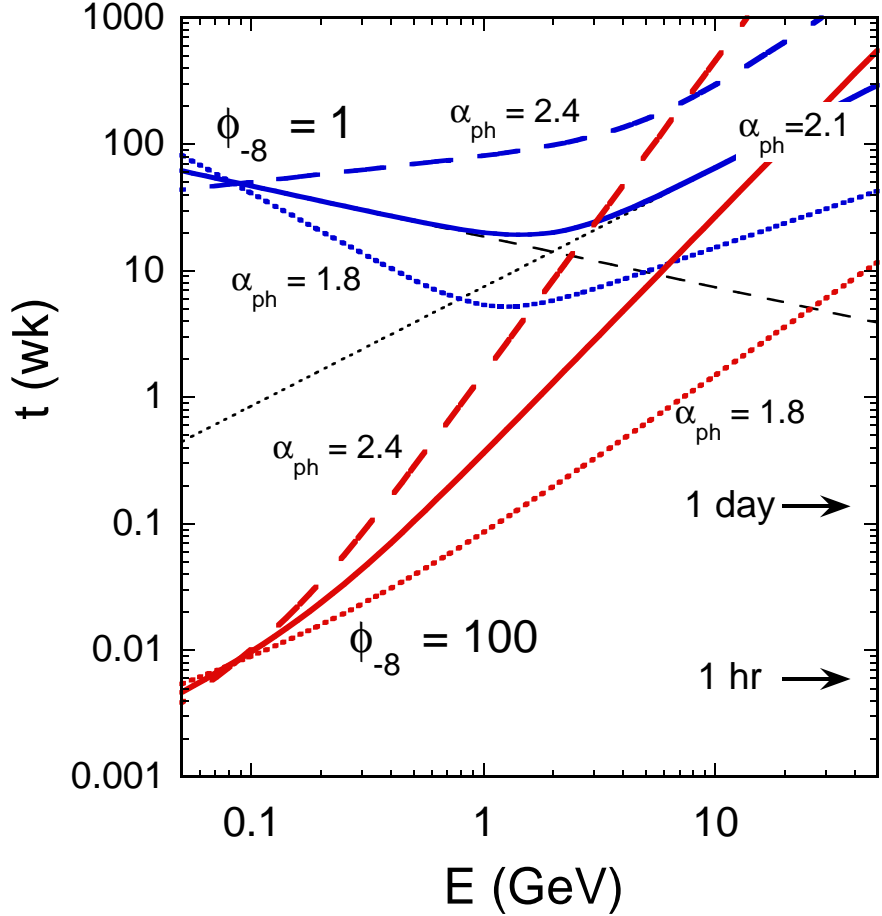}
\vskip-0.1in
\caption{Time required for GLAST in the scanning mode to detect a 
source at integral flux levels $\phi_{-8} =1$ and $\phi_{-8} =100$
when integrating signal above energy $E$, for sources with photon
spectral indices $\alpha_{ph} = 2.4$ (thick long-dashed curves),
$\alpha_{ph} = 2.1$ (thick solid curves), and $\alpha_{ph} = 1.8$
(thick dotted curves).  Light dotted and dashed lines show the time
required to detect a source from signal and background limits,
respectively, for $\phi_{-8} =1$ and $\alpha_{ph} = 2.1$. }
\label{fig9}
\end{figure}

A better analysis considers optimal energy for source detection from
both signal and noise limits.  Equating eqs.\ (\ref{phim8G}) and
(\ref{twkg}) for the minimum time needed to detect sources at the
5$\sigma$, 5 count limit assuming $p = 3.2$ or $\alpha_\nu = -0.1$,
gives the best lower photon energy above which to integrate, namely
\begin{equation}
\bar u \cong 20\;\phi_{-8}^{-3/4}\;,\;
\;{\rm or}\;\;\bar E \cong 2.0 \phi_{-8}^{-3/4}\;{\rm GeV}\;.
\label{baru}
\end{equation}
The time it takes to reach this flux level is
\begin{equation}
\bar t_{wk}\cong {14\over \phi_{-8}^{1.7}}\;.
\label{bartwk}
\end{equation}
Fig.\ 12 shows the minimum time for GLAST to detect sources at
different integral flux levels $\phi_{-8}$ and with different spectral
indices from the signal and background limits, eqs.\ (\ref{phim8G})
and (\ref{twkg}), respectively.  In order to determine the source's
spectral index, a hardness ratio or mean photon energy $\langle E
\rangle $ can be calculated for photons originating from a source at a
known source position. For photons with energies between $100$ MeV and
$E_2 = 100 u_2 $ MeV originating from a source with spectral index
$\alpha_{ph}$,
\begin{equation}
\langle E \rangle \;=\; E_{100} 
\big( {\alpha_{ph}- 1 - a_0\over \alpha_{ph} - 2 - a_0}\big)\;
{1-u_2^{\alpha_{ph} - 2 - a_0} \over 1 - u_2^{\alpha_{ph} - 1 - a_0}}\;.
\label{averagee}
\end{equation} 
Background contamination must also be considered to determine the best
value and error for $\alpha_{ph}$. The upper energy used for $u_2$
depends on the nature of the source spectrum or the cutoff produced by
$\gamma\gamma$ attenuation on the EBL. For low redshift sources,
contemporaneous observations with MAGIC, HESS, or VERITAS ground-based
$\gamma$-ray telescopes could help constrain the upper energy. Higher
energy analysis can enhance blazar detection. For example,
\citet{db01} reported the search for blazars in the EGRET data by
looking at $> 10$ GeV photons.

From the previous considerations and from Fig.\ 9, weak sources are best
identified with GLAST by calculating the statistical significance $S(>E)/\sqrt{B(>E)}$ 
of a source 
as a function of photon energy $E$, and searching for the minimum in the 
detection significance. The actual measurement consists of the sum $S(>E)$ and
$B(>E)$, and the value of $B(>E)$ is calculated from our knowledge
of the diffuse galactic and extragalactic background and GLAST detection characteristics.
Knowledge of $B(>E)$ will improve with our understanding of the GLAST results. 
For sources with spectra harder than the diffuse EGRB, a minimum in the 
combined background and signal-limited significance will be found by this procedure.
This represents a systematic bias for GLAST to discover hard spectrum sources, and must
also be considered when treating population statistics. At the same time, an unbiased
search for sources with fluxes above a fixed energy $E$, e.g., $E = 100$ MeV, $1 $ GeV, 
and 10 GeV should also be made with GLAST.

\subsection{Short Timescale Flares}

GLAST reaches EGRET 2-week sensitivities of $\phi_{-8}\cong 15$ after
$\approx 1$ -- 2 days, as can be obtained by comparing eq.\
(\ref{bartwk}) with eq.\ (\ref{twkg}).  It reaches this sensitivity
over the {\it full sky}, compared to $\approx 1/24^{th}$ of the full
sky viewed with EGRET. For a mean photon energy of $\approx 400$ MeV
measured with EGRET from a blazar with $2\lesssim \alpha_{ph}
\lesssim 2.5$, a measurement of $\phi_{-8}\cong 15$ (150)
corresponds to an energy flux between 100 MeV and 5 GeV of $\cong
10^{-10}$ $(10^{-9}$) ergs cm$^{-2}$ s$^{-1}$.

Consider a flare at the level of $\phi_{-8}\cong 200$, corresponding
to some of the brightest flares seen with EGRET. Given the Phase 1
EGRET all-sky survey results \citep{dd03}, we predict a flare at this
level every few days, with large uncertainties given the small
statistics of bright blazar flares. For a blazar with $\gamma$-ray
spectral index $\alpha_{ph} = 2.1$, GLAST will have a significant
$5\sigma$ detection after
\begin{equation}
\bar t \cong 2.6 \;{\rm ks}\;,
\label{twk}
\end{equation}
using $\bar u = 1$ whenever $\bar u < 1$. Better sensitivity of a
bright blazar flare is possible for observing times less than the
GLAST orbital period of 1.6 hr (or 5.8 ks) if the blazar is in the FOV
of GLAST. In this case, $X_s$ can be larger than 1/5.  These
timescales are of particular interest inasmuch as we expect a minimum
variability timescale $t_{var}$ corresponding to a light-crossing
distance
\begin{equation}
t_{var} = {2GM\over c^3}\;(1+z) \approx M_8(1+z)\;{\rm ks}\;;
\label{cDeltatvar}
\end{equation}
corrections should be made to this expression for a Kerr metric.
Temporal power analyses of strings of data from bright blazars, such
as 3C 279 and PKS 0528+134, could find a reduction of power at a
timescale corresponding to the Schwarzschild radius or radius of the
minimum stable orbit.  Rare, bright, hard spectrum flares are of
particular interest for GLAST analysis in order to look for
$\gamma\gamma$ absorption cutoffs, whether internal (within the blob),
external within the inner jet or galactic environment, or due to
absorption on diffuse radiation fields.

\section{Synchrotron Self-Absorption Frequency}

We calculate the synchrotron self-absorption (SSA) frequency
in terms of the parameters of the system, $l^\prime_e, \Gamma,$ and  $z$. Using
the formalism of \citet{gou79}, cast into dimensionless notation, 
we have the expression for the frequency $\nu_m$ where
$d F_\nu(\nu_m)/d\nu = 0$, given by
\begin{equation}
\e_m = {h\nu_m\over m_ec^2} = {\delta_{\rm D}\over 1+z}\;\big[{4\pi c(p)\over \alpha_f t_m}
r_b^\prime r_e^2 k_e^\prime\big]^{2/(p+4)} b^{(p+2)/(p+4)}\;,
\label{em}
\end{equation}
and $c(p)$ and $t_m$ are numbers of order unity \citep{gou79}, $\alpha_f = 1/137$
is the fine structure constant, and $K_e^\prime = 
V_b^\prime k_e^\prime$. From the expression 
for synchrotron radiation, eq.\ (\ref{leprime5}),
\begin{equation}
l_e^\prime = \big({2c\sigma_{\rm T} U_{B_{cr}}\over 9}\big) \; k_e^\prime r_b^{\prime ~3}\;b^{(p+1)/2}\;.
\label{leprime1}
\end{equation}
It follows from these expressions
\begin{equation}
\e_m ={\nu({\rm GHz})\over 1.2\times 10^{11}}
= 
{\delta_{\rm D}\over 1+z}\;\big[{18\pi c(p)\over \alpha_f t_m}
{l_e^\prime \over c \sigma_{\rm T} U_{B_{cr}} }\;({r_e\over r_b^\prime}\big)^2\big]^{3/(p+4)} \; b^{1/(p+4)}\;.
\label{em1}
\end{equation}
We obtain
\begin{equation}
\nu({\rm GHz})\cong  
{\delta_{\rm D}\over 1+z}\;
\cases{\;0.22 \;\big(\sqrt{l_{42}}/ r_{14}\big)\;B({\rm G})^{1/6}\; \; , & 
$p=2$\cr\cr 9.5 \;\big({l_{42}/ r^2_{14}}\big)^{3/7}\;B({\rm G})^{1/7}\;,
 & $p=3$\cr\cr 1900 \;\big({l_{42}/ r^2_{14}}\big)^{3/8}\;B({\rm G})^{1/8}\;,
 & $p=4$\cr}\;,
\label{num1}
\end{equation}
where $r_b^\prime = 10^{14} r_{14}$ cm and $l_e^\prime = 10^{42} l_{42}$
ergs s$^{-1}$ sr$^{-1}$.
 BLs are likely to 
be heavily self-absorbed at 5 GHz, using
 parameters  $\delta_{\rm D}\cong 4$, 
$p = 3$, $l_{42} \cong 1$, $r_{14} \cong 1$,
and $B\cong 0.1$ -- 1 G.
FSRQs could also be self-absorbed at 5 GHz, using
 $\delta_{\rm D}\cong 10$, $p > 3$, $l_{42} \cong 0.01$,
$r_{14} \cong 100$, and $B \gtrsim 1$ G. Here the values 
of $B$ are typical of equipartition
magnetic field strengths. If GLAST shows 
that FSRQs vary on hour or subh-our timescales, then 
the synchrotron emission of blazars is likely to be strongly 
self-absorbed.


\begin{deluxetable}{lccccc}
\tablecaption{Sample of High-Confidence Gamma-Ray Blazars used in Analysis\label{table1}}
\tablehead{   Catalog Name  &
\multicolumn{1}{c}{$\phi^{pk}_{-8}$\tablenotemark{a}} ~~& 
\multicolumn{1}{c}{$\Delta \phi^{pk}_{-8}$}~~&
~~Redshift $z$~~& 
\multicolumn{1}{c}{Other Name\tablenotemark{b}  } ~~ &
  \multicolumn{1}{c}{Classification\tablenotemark{c}}
}
\startdata
\tableline
\tableline
3EG J0204+1458 & 	 52.8 & 	26.4 & 	 0.405 & PKS 0202+14\\	
3EG J0210-5055  & 	 134.1 & 	24.9 & 	 1.003 & PKS 0208-512\\	
3EG J0222+4253 & 	 25.3 & 	5.80 & 	 0.444 & 	3C 66A   &		 B\\
3EG J0237+1635 & 	 65.1 & 	8.80 & 	 0.940 & 	AO 0235+164 & 	 B\\
3EG J0340-0201 & 	 177.6 & 	36.6 & 	 0.852 & PKS 0336-01	\\
3EG J0412-1853 & 	 49.5 & 	16.1 & 	 1.536 & PKS 0414-189\\	
3EG J0422-0102 & 	 64.2 & 	34.2 & 	 0.915 & PKS 0420-01\\	
3EG J0442-0033 & 	 85.9 & 	12.0 & 	 0.844 & 	NRAO 190\\      
3EG J0450+1105 & 	 109.5 & 	19.4 & 	 1.207 & 	PKS 0446+112\\
3EG J0456-2338 & 	 14.7	 & 4.20 & 	 1.009 & 	PKS 0454-234\\
3EG J0458-4635 & 	 22.8 & 	7.40 & 	 0.8580 &  PKS 0454-46	\\
3EG J0459+0544 & 	 34.0 & 	18.0 & 	 1.106 & PKS 0459+060	\\
3EG J0500-0159 & 	 68.2 & 	41.3 & 	 2.286 & PKS 0458-02	\\
3EG J0530+1323 & 	 351.4 & 	36.8 & 	 2.060 & 	PKS 0528+134\\
3EG J0540-4402 & 	 91.1 & 	14.6 & 	 0.894 &    PKS 0537-441 & 	 B\\
3EG J0721+7120 & 	 45.7 & 	11.1 & 	 0.30 & S5	0716+714 & 		 B\\
3EG J0737+1721 & 	 29.3 & 	9.90 & 	 0.424	 & PKS 0735+178		&	B\\
3EG J0743+5447 & 	 42.1 & 	8.30 & 	 0.723& RX J0742.6+5444	\\
3EG J0828+0508 & 	 35.5 & 	16.3 & 	 0.180 & PKS 0829+046	 & 			 B\\
3EG J0829+2413 & 	 111.0 & 	60.1 & 	 0.939 & OJ 248	\\
3EG J0845+7049 & 	 33.4 & 	9.00 & 	 2.172 & 4C 71.07	\\
3EG J0852-1216 & 	 44.4 & 	11.6 & 	 0.566 & PMN J0850-1213	\\
3EG J0853+1941 & 	 15.8 & 	6.90 & 	 0.306 & 	OJ 287	 & 	 B\\
3EG J0952+5501 & 	 47.2 & 	15.5 & 	 0.901 & OK 591	\\
3EG J0958+6533 & 	 18.0 & 	9.40 & 	 0.368 &  S4 0954+65 & 				 B\\
3EG J1104+3809 & 	 27.1 & 	6.90 & 	 0.031 & 	Mkn 421 &  	 B\\
3EG J1200+2847 & 	 163.2 & 	40.7	 &  0.729 & TON 0599  & 	 \\
3EG J1222+2841 & 	 53.6 & 	14.1 & 	 0.102 & 	W Comae & 		 B\\
3EG J1224+2118 & 	 48.1 & 	15.3 & 	 0.435 & PG 1222+216	\\
3EG J1229+0210 & 	 48.3 & 	11.3 & 	 0.158 & 	3C 273\\
3EG J1230-0247 & 	 15.5	 & 4.10 & 	 1.045 & PKS 1229-02	\\
3EG J1246-0651 & 	 44.1 & 	29.6 & 	 1.286 & PKS 1243-072	\\
3EG J1255-0549 & 	 267.3 & 	10.7 & 	 0.538 & 	3C 279\\
3EG J1329+1708 & 	 33.1 & 	19.3 & 	 2.084 & OP 151	\\
3EG J1339-1419 & 	 20.2 & 	11.6 & 	 0.539 & PKS 1335-127	\\
3EG J1409-0745 & 	 128.4 & 	23.4 & 	 1.494 & PKS 1406-076	\\
3EG J1429-4217 & 	 	 55.3	 & 16.3	 &  1.522 & PKS 1424-41	\\
3EG J1512-0849 & 	 49.4 & 	18.3 & 	 0.361 & PKS 1510-08	\\
3EG J1605+1553 & 	 42.0 & 	12.3 & 	 0.357 & 4C 15.54 & 	 B\\
3EG J1608+1055 & 	 62.4 & 	13.0	  & 1.226 & OS 111	\\
3EG J1614+3424 & 	 68.9 & 	15.3 & 	 1.401 & OS 319 	\\
3EG J1625-2955 & 	 258.9 & 	15.3 & 	 0.815 & PKS 1622-29\\	
3EG J1626-2519 & 	 82.5 & 	35.0 & 	 0.786 & PKS 1622-253	\\



3EG J1635+3813 & 	 107.5 & 	9.60 & 	 1.814 & 4C 38.41\\	
3EG J1727+0429 & 	 30.2 & 	18.8 & 	 0.296 & PKS 1725+044	\\
3EG J1733-1313 & 	 104.8 & 	34.7 & 	 0.902 & PKS 1730-13 \\	
3EG J1738+5203 & 	 44.9	 & 26.9 & 	 1.375 & OT 566	\\
3EG J1744-0310 & 	 48.7	 & 19.6 & 	 1.054 & PKS 1741-03	\\
3EG J1935-4022 & 	 93.9 & 	31.4 & 	 0.966 & PKS 1933-400	\\
3EG J1937-1529 & 	 55.0 & 	18.6	  & 1.657 & PKS 1936-15	\\
3EG J2025-0744 & 	 74.5 & 	13.4	 &  1.388 & PKS 2023-07	\\
3EG J2036+1132 & 	 35.9	 & 15.0 & 	 0.601 & TXS 2032+117 &  B\\
3EG J2055-4716 & 	 35.0	 & 20.9	 &  1.489 & 	PKS 2052-47 \\
3EG J2158-3023 & 	 30.4 & 	7.70 & 	 0.116 & 	PKS 2155-304 & B\\
3EG J2202+4217 & 	 39.9 & 	11.6 & 	 0.069 & 	BL Lac &	 B\\
3EG J2232+1147 & 	 51.6 & 	15.0 & 	 1.037 & 	CTA 102\\
3EG J2254+1601 & 	 116.1 & 	18.4 & 	 0.859 & 	3C 454.3\\
3EG J2321-0328 & 	 38.2 & 	10.1 & 	 1.411 & PKS 2320-035	\\
3EG J2358+4604 & 	 42.8 & 	20.3	 &  1.992 & OZ 486	\\
3EG J2359+2041 & 	 26.3 & 	9.00 & 	 1.066 & OZ 193\\
\tableline
\enddata
\tablenotetext{a}{$\phi^{pk}_{-8}$: peak flux ($E > 100$ MeV)  in units of $10^{-8}$ photons cm$^{-2}$ s$^{-1}$.}
\tablenotetext{b}{Survey catalogs include PKS: Parkes, O(A-Z): Ohio, 3C/4C: Cambridge, and
TXS: Texas.} 
\tablenotetext{c}{B: BL Lac object; no entry: FSRQ.} 
\end{deluxetable}

\end{document}